\documentclass[letterpaper]{JHEP3}

\usepackage{epsfig}
\usepackage{graphicx}

\newcommand{\labell}[1]{\label{#1}}  
\newcommand{\reef}[1]{(\ref{#1})}
\newcommand{\ie}{{\it i.e.,}\ }
\newcommand{\eg}{{\it e.g.,}\ }
\newcommand{\ssc}{\scriptscriptstyle}
\newcommand{\GN}{G_{\ssc N}}
\newcommand{\rr}{\rho}
\def\IR{{\hbox{{\rm I}\kern-.2em\hbox{\rm R}}}}
\newcommand{\be}{\begin{equation}}
\newcommand{\ee}{\end{equation}}
\newcommand{\beq}{\begin{equation}}
\newcommand{\eeq}{\end{equation}}
\newcommand{\prt}{\partial}
\newcommand{\Lam}{\Lambda}
\newcommand{\lc}{\ell}

\preprint{{\small hep-th/0202094}}


\title{Tall tales from de Sitter space I:\\Renormalization group flows}
\author{Fr\' ed\' eric Leblond\footnote{E-mail: {\tt fleblond@hep.physics.mcgill.ca}},
Donald Marolf\footnote{E-mail: {\tt marolf@physics.syr.edu}} and
Robert C. Myers\footnote{E-mail: {\tt rcm@hep.physics.mcgill.ca}}\\
$^{*,\ddagger}$ Department of Physics, McGill University, Montr\' eal, Qu\' ebec H3A 2T8\\
$^\dagger$ Physics Department, Syracuse University, Syracuse, New York 13244\\
$^\ddagger$ Perimeter Institute for Theoretical Physics, Waterloo, Ontario N2J 2W9\\
$^\ddagger$ Department of Physics, University of Waterloo, Waterloo, Ontario N2L 3G1}
\date{December, 2001}
\abstract{We study solutions of Einstein gravity coupled to a positive
cosmological constant and matter which are asymptotically de Sitter and homogeneous.
Regarded as perturbations
of de Sitter space, a theorem of Gao and Wald implies that generically
these solutions are `tall,' meaning that the
perturbed universe lives through enough conformal time for an entire
spherical Cauchy surface to enter any observer's past light cone.
Such observers will realize that their universe is spatially compact.
An interesting fact, which we demonstrate with an explicit example, is that this
Cauchy surface can have arbitrarily large volume for fixed
asymptotically de Sitter behavior.
Our main focus is on the implications of tall universes for the
proposed dS/CFT correspondence. Particular attention is given to the associated
renormalization group flows, leading to a more general de Sitter `c-theorem.'
We find, as expected,  that
a contracting phase always represents a flow toward the infrared,
while an expanding phase represents a `reverse' flow toward the ultraviolet.
We also discuss the conformal diagrams for various classes of homogeneous
flows.}

\begin{document}
\section{Introduction}
\label{intro}

\setcounter{footnote}{0}

Recent observations suggest that our universe is proceeding toward a
phase where its evolution will be dominated by a small positive cosmological
constant --- see, \eg \cite{supernova}. These results increase the
urgency with which physicists have addressed the question of understanding
the physics of de Sitter-like spacetimes --- see, \eg
\cite{AS1,AS,BBM,witten,AS3,marcus,Banks,BHM,robb,test,klemm}.
While de Sitter (dS) space does not itself represent a phenomenologically interesting
cosmology, it does present a simple framework within which we may investigate
the physics of quantum gravity with a positive cosmological constant.
In particular, the cosmological horizon of de Sitter space is
an interesting and oft-discussed feature which
appears in many spacetimes having a positive cosmological constant.
A recent development has been the conjecture \cite{AS1} that quantum gravity
in such spacetimes has a dual description in terms of a Euclidean conformal
field theory on the future boundary ($I^+$)
and/or the past boundary ($I^-$). These ideas are naturally extended to include
general solutions of Einstein gravity coupled to a positive cosmological constant with
asymptotically de Sitter regions to the past and/or future. The time evolution
of these solutions corresponds to a re-scaling of the boundary metric and so
within the context of the dS/CFT duality, the evolution has a natural interpretation
in terms of a renormalization group flow \cite{AS,BBM}. Similar renormalization
group flows have also been discussed in the context of stringy cosmologies
\cite{miao}.

\FIGURE{\epsfig{file=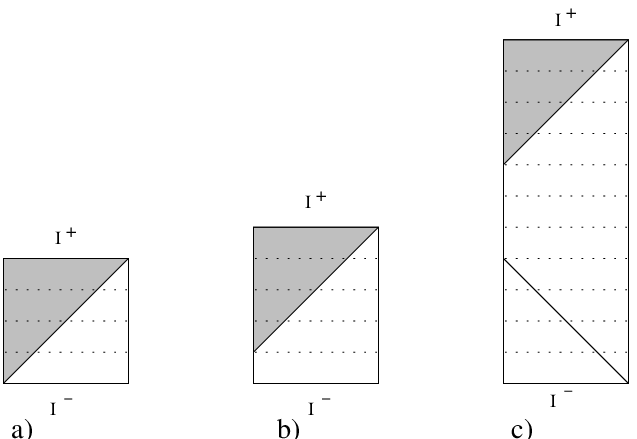}\caption{Conformal diagrams
of a) de Sitter space, b) perturbed de Sitter
space, and c) a very tall asymptotically de Sitter spacetime. The worldline of
the `central observer' is the right boundary of each diagram and various
horizons related to her worldline are shown.  Shaded regions cannot send
signals to this observer.} \label{heights}}

An interesting property of de Sitter space is that it has
compact Cauchy surfaces.\footnote{As we will be considering dS spaces
of arbitrary dimensions, we should add that this statement need not apply in
two spacetime dimensions, where the universal cover of dS space
has an infinite spatial volume.} However, as illustrated in
figure \ref{heights}(a), the causal structure of pure de Sitter space is such
that an observer can never see an entire compact Cauchy
surface. Instead, the observer's light cone only expands to include the
full Cauchy surface at $I^-$, asymptotically as the observer approaches
$I^+$ . This causal connection between $I^+$ and $I^-$ plays an important
role in understanding the role of both of these surfaces in the dS/CFT
correspondence \cite{AS1,witten,AS3,marcus}. However, a theorem (Corollary 1)
of Gao and Wald \cite{GW} tells us that under a
generic perturbation\footnote{Technically, any perturbation
satisfying the null generic condition \cite{Wald}.}  of de Sitter
space, the conformal diagram becomes taller so that an entire
compact Cauchy surface now becomes visible at some finite time.
This is shown in figure \ref{heights}(b).   Pushing this somewhat
further, one can imagine that in certain circumstances asymptotically de
Sitter spacetimes of the sort shown in figure \ref{heights}(c) may arise.
That is, in these spacetimes, a compact Cauchy surface lies in the
intersection of the past {\it and} future of a generic worldline. In fact,
we construct explicit examples of such spacetimes in section \ref{construct}.
These particular examples are of some interest with regard to general
discussions of `the number of degrees of freedom' in asymptotically de Sitter
spaces \cite{Banks,Nbound,hunt}, as the region open to experimental
probing by an observer contains arbitrarily large spatial volumes.

The primary focus of the present paper is investigating
renormalization group flows in the context of the dS/CFT duality.
The paper is organized as follows: In section \ref{fol},
we review the basics of de Sitter space, establish our conventions,
and introduce the three types of foliations (spherical, flat, and hyperbolic)
that will be most relevant in the following sections. Section
\ref{gass} then generalizes to include a rolling scalar field,
which may yield `flows' which are asymptotically dS.
In particular, we describe two solution generating techniques
which may be used to construct explicit solutions.
Section \ref{ct} is devoted to extending the `c-theorem' for asymptotically
de Sitter evolutions \cite{AS,BBM}. We generalize the `c-theorem' to include
flows with spherical or hyperbolic spatial sections and also certain situations
where the spatial geometries are anisotropic. We find that under these general
circumstances, an `effective' cosmological constant always decreases (increases)
during an expanding (contracting) phase. Finally we consider the possibilities for
transitions between phases of expansion and contraction. Because of our interest
in global structures, we discuss the Penrose diagrams
relevant to these flows in section \ref{diagrams}.
Section \ref{construct} describes the construction of a `very tall'
universe.  Within the context of
one model, we illustrate solutions with asymptotically de Sitter
regions both to the future and past and that contain an arbitrarily
long lived intermediate matter dominated phase in which the
spatial volume is arbitrarily large.
We conclude with a discussion of our results in section \ref{disc}.
Finally, with three appendices, we provide additional examples of asymptotically
de Sitter renormalization group flows, and flesh out the construction of the
conformal diagrams discussed in section \ref{diagrams}.

\section{De Sitter space basics}\label{fol}

The simplest construction of the ($n$+1)-dimensional de Sitter (dS) spacetime is
through an embedding in Minkowski space in $n+2$ dimensions, where it may
be defined as the hyperboloid
\be
\labell{hyper}
\eta_{AB}X^AX^B=\lc^2.
\ee
The resulting surface is maximally symmetric, {\it i.e.,}
\be
\labell{maxi}
R_{ijkl}={1\over \lc^2}(g_{ik}\,g_{jl}-g_{il}\,g_{jk})\ ,
\ee
which also ensures that the geometry is locally conformally flat.
Hence dS space solves Einstein's equations with a positive
cosmological constant,
\be
\labell{einst}
R_{ij}={2\Lam\over n-1}g_{ij}\qquad{\rm with}\quad\Lam={n(n-1)\over2\lc^2}\ .
\ee
The topology of the space is $R\times S^n$. The Penrose
diagram is represented by a square \cite{HawEll}, as illustrated in Figure
\ref{heights} (a).
Any horizontal cross section of the figure is an $n$-sphere, so that any point
in the interior of the diagram represents an $(n-1)$-sphere. At the right and left edges,
the points correspond to the north and south poles of the $n$-sphere. Note that
diagram is just tall enough that a null cone emerging from, say, the south pole
at $I^-$ reconverges on the north pole at $I^+$.

One may present the metric on dS space in many different coordinate systems,
which may be particularly useful in different situations. Three
simple  choices for ($n$+1)-dimensional dS space come from foliating
the embedding space above with flat hypersurfaces, $n_AX^A=constant$.
The three distinct choices correspond to the cases where the normal
vector $n_A$ is time-like, null or space-like. With these distinct
choices, a given hypersurface intersects the hyperboloid on a spatial
section which has a spherical, flat or hyperbolic geometry, respectively.
Following the standard notation for Friedmann-Robertson-Walker cosmologies,
we denote these three cases as $k=+1,0$ and $-1$, respectively.
Then the three corresponding metrics on dS space
can be written in a unified way as follows:
\be
\labell{metric2b}
ds^2=-dt^2+a_k^2(t) d\Sigma^2_{k,n}\ ,
\ee
where the $n$-dimensional Euclidean metric $d\Sigma^2_{k,n}$ is
\begin{equation}
\labell{little}
d\Sigma^2_{k,n} =\left\{ \begin{array}{ll}
\vphantom{\sum_{i=1}^{n}}
\lc^2d\Omega^2_{n}& {\rm for}\; k = +1\\
\sum_{i=1}^{n} dx_i^2&{\rm for}\; k = 0 \\
\vphantom{\sum_{i=1}^{n}}
\lc^2d\Xi^2_{n} &{\rm for}\; k = -1\ ,
\end{array} \right.
\end{equation}
where $d\Omega^2_{n}$ is the unit metric on $S^{n}$. The
`unit metric' $d\Xi^2_{n}$ is
the $n$--dimensional hyperbolic space ($H^{n}$)
which can be obtained by analytic continuation of
that on $S^n$.  For $k{=}\pm 1$ we assume that $n\ge 2$.

The scale factor in each of these cases would be given by
\begin{equation}
\labell{radd}
a_{k}(t) =\left\{ \begin{array}{lll}
\cosh (t/\lc)& {\rm for}\; k = +1\\
\exp({t/\lc})& {\rm for}\; k = 0 \\
\sinh(t/\lc)&{\rm for}\; k = -1\ .
\end{array} \right.
\end{equation}
Hence we see that $k=+1$ corresponds to the standard global coordinates, in
which the spatial $n$-sphere begins by shrinking from infinity to a minimum
size (with radius $\lc$) and then it re-expands. The choice $k=0$ corresponds to the standard
inflationary coordinates, where the flat spatial slices experience an
exponential expansion (assuming a positive sign in the exponential).
In this case, $t=-\infty$ corresponds to a horizon (\ie the boundary
of the causal future) for a co-moving observer emerging from $I^-$. Hence these coordinates only
cover half of the full dS space but, of course, substituting a minus sign in the
exponential of eq.~\reef{radd} yields a metric which naturally covers the lower half.
The choice $k=-1$ yields a perhaps less familiar
coordinate choice where the spatial sections have constant negative curvature.
In this case, $t=0$ again represents a horizon.  However,
this horizon is the future null cone of an actual point inside dS
space. Figure \ref{slices}
illustrates slices of constant $t$ on a conformal diagram of dS space.
It is straightforward to see that all of the above
coordinate systems in fact are related by a local diffeomorphism.
One might also note the exponential expansion that dominates the late
time evolution of all three slicings independent of the spatial curvature,
\ie $a_k(t)\sim exp(t/\lc)$ as $t\rightarrow\infty$ for all $k$.

These three different coordinate patches are displayed on the full Penrose
diagram in figure \ref{slices}. One comment on the $k=-1$ case is that
the central diamond cannot be foliated by homogeneous spacelike hyperboloids.
However, it is naturally foliated by timelike hyperboloids, \ie by
copies of dS space.  The metric in the central diamond region may
be obtained by double analytic continuation of \reef{metric2b} and the result is
\begin{equation}
\labell{tlh}
ds^2  = dt^2 + \lc^2\sin^2(t/\lc) ds^2_{dS},
\end{equation}
where $ds^2_{dS}$ is the metric for $n$-dimensional de Sitter space
with unit radius of curvature.
Here $t$ is a spacelike coordinate that is naturally thought of as the
analytic continuation of the $t$ in (\ref{radd}) behind the horizon.

\FIGURE{\epsfig{file=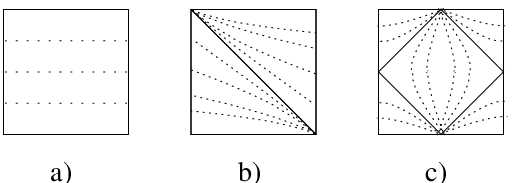, width = 7cm}\caption{Constant $t$ slices in a)
the spherical slicing, b) the
flat slicing, and c) the hyperbolic slicing of de Sitter space.}\label{slices}}

With the metric~\reef{metric2b}, we have apparently displayed dS$_{n+1}$ with
three different boundary geometries:
\be
\labell{display}
S^{n},\quad \IR^n\quad {\rm and}\quad H^n\ .
\ee
The dS/CFT correspondence would imply then an equivalence between, on the
one hand, quantum gravity in dS space and a CFT on any
of the above backgrounds \reef{display}. On the other hand, we know that
the (past and future) timelike infinities of dS$_{n+1}$ space have the topology
$S^n$. Hence the correct observation is that the latter two boundary geometries
are equivalent to (a portion of) $S^n$ up to a conformal transformation.
However, a singularity in the latter transformation
effectively removes (\ie pushes off to infinity) certain points on the sphere
to produce the resulting geometry. In the case where the boundary appears to be
$H^n$, it is obvious from the Penrose diagram (see figure
\ref{slices}) that these coordinates cover only  half of the boundary of dS space.
Hence the conformal transformation has pushed out the equator of the boundary
sphere to produce two copies of the hyperbolic plane.

A common feature of the three coordinate systems \reef{metric2b} above
is that the spatial geometry is uniformly scaled in the time evolution.
In appendix \ref{folb}, we present some other metrics on dS space which
have non-uniform scalings. In particular, the boundary metric takes the
form of a direct product of two submanifolds and each of the latter
submanifolds evolves with a different scale factor.

\section{Generating asymptotically de Sitter solutions}
\label{gass}

In this section, we set forth a framework in which to
consider generalized dS flows or asymptotically dS solutions.
In investigating new phenomena, it is useful to have a set of
explicit solutions at one's disposal.  Hence, we introduce
two solution generating techniques below.  Section \ref{exactf}
describes a method that allows for a variety of potentials but is
useful mainly for the case of spatially flat sections ($k=0$).
This approach is a simple adaptation of the techniques developed in
\cite{gub1} for the case of spatially varying solutions with a negative
cosmological constant.
This approach allows the equations of motion to be reduced to two
first order ODE's.
Section \ref{step} treats the special case of piece-wise constant potentials in
a way that remains useful for any value of $k$.  Throughout this section,
our attention is restricted to spacetimes with homogeneous spatial sections.

In general we consider $(n+1)$-dimensional models of Einstein gravity coupled
to a scalar field $\phi$. We write the action as
\beq
\labell{totact}
S = \frac{1}{16\pi \GN}\int d^{n+1}x \sqrt{-g}\,\left[R
- n(n-1)g^{ij}\partial_{i}\phi \partial_{j}\phi-n(n-1) V(\phi)\,\right]\ .
\eeq
Note that the scalar field terms have been normalized in
an unconventional manner (including the fact that Newton's constant $\GN$ appears in an
overall factor in front of the total action) to simplify Einstein's equations in the following
analysis. Einstein's equations may be written as
\beq
\labell{eomf}
R_{ij} -\frac{1}{2}g_{ij}R = T_{ij},
\eeq
where the stress-energy tensor is given (with a slightly unconventional normalization) as
\beq
T_{ij} = n(n-1)\left[ \partial_{i}\phi\partial_{j}\phi
-\frac{1}{2}g_{ij}\left(\,(\partial\phi)^2+V(\phi)\,\right)\,\right]\ .
\eeq
Note that if the scalar field sits at a critical point $\phi=\phi_0$ of the potential $V(\phi)$,
the effective cosmological constant is given by $\Lambda=\frac{n(n-1)}{2}V(\phi_0)$.

Now we will consider the spatially homogeneous solutions of the form
\be
\labell{ansatz2c}
\phi=\phi(t)\quad{\rm and}\quad ds^2=-dt^2+a^2(t) d\Sigma^2_{k,n}
\ee
with $d\Sigma^2_{k,n}$ defined in eq.~\reef{little}.
Given this ansatz, the scalar field equation reduces to
\begin{equation}
\labell{e3}
\ddot{\phi} + n \frac{\dot{a}}{a} \dot{\phi} = -\frac{1}{2}\frac{\partial V}{\partial \phi}\ ,
\end{equation}
where a `dot' denotes a derivative with respect to $t$.
The dynamics of the scale factor $a(t)$ is governed by the Friedmann equations
\begin{equation}
\labell{e1}
\left( \frac{\dot{a}}{a} \right)^{2} + \frac{k}{a^2} = \dot{\phi}^{2} + V(\phi)\ ,
\end{equation}
\begin{equation}
\labell{e2}
\frac{\ddot{a}}{a} = -(n-1)\dot{\phi}^{2} + V(\phi)\ .
\end{equation}
The second of these is redundant, and a complete solution may be determined from
eqs.~\reef{e3} and \reef{e1} alone.

\subsection{Pre-potentials}
\label{exactf}

One approach to producing explicit solutions is
to adapt the technique of \cite{gub1} to the present case.
This method considers potentials of a special form which
allow us to simplify the equations of motion, \reef{e3} and \reef{e1},
for both $a(t)$ and $\phi(t)$ to a system of two first order ODE's.
It is easily verified that eqs.~\reef{e3}, (\ref{e1}) and (\ref{e2})
are satisfied when $\phi$, $\dot{a}/a$, and $V(\phi)$ are related to a pre-potential
$W(\phi)$ through
\beq
\labell{sysfaa}
\dot{\phi} = \frac{1}{\gamma} \frac{\partial W(\phi)}{\partial \phi} \;\;\;\;\;\;
\frac{\dot{a}}{a}=-\gamma n W(\phi),
\eeq
\beq
\labell{potf2}
V(\phi) = -\frac{1}{\gamma^{2}}\left(\frac{\partial W(\phi)}{\partial \phi}\right)^{2}
+ n^{2} W^{2}(\phi),
\eeq
where
\beq
\gamma = \left( 1 - \frac{k}{n^{2}W(\phi)^{2}a^{2}}\right)^{1/2}.
\eeq
For $k=\pm1$, these expressions are of limited use as they are highly nonlinear
and further the scale factor $a(t)$ appears
in eq.~\reef{potf2} through the factor of $\gamma^2$. However, for $k=0$, $\gamma=1$ and
the scalar potential is completely determined by the pre-potential $W(\phi)$.
In this case, the equations \reef{sysfaa} reduce to
\beq
\labell{sysfs}
\dot{\phi} = \frac{\partial W(\phi)}{\partial \phi} \;\;\;\;\;\;
\frac{\dot{a}}{a} =-n W(\phi)\ ,
\eeq
which can readily be solved analytically given a sufficiently simple $W(\phi)$.
While this technique allows us to construct many interesting analytic flows,
we will not consider any explicit examples here. 

\subsection{Step potentials}
\label{step}

Another class of tractable models arises when $V(\phi)$ is piece-wise
constant.  On each constant piece, the system can be mapped to a particle
in a one-dimensional potential so that the dynamics is conveniently
summarized by the effective potential. One may then
patch together the solutions at the boundaries of the steps.
With some number of small steps, this approach may be useful to
approximate a slowly varying potential. Alternatively, this technique
may be used to simulate the effect of phase transitions
in the matter sector. We will consider an interesting example based
on these models in section \ref{construct}.

Let us first consider the dynamics within one step with constant
$V(\phi)=V_0$, which is assumed to be positive. Recall that in this phase
of the evolution, the cosmological constant is
given by $\Lambda=n(n-1)V_0/2$. From (\ref{e3}), the scalar field equation reduces to
\begin{equation}
\labell{phieq0}
0= \ddot{\phi} + n \frac{\dot{a}}{a} \dot{\phi} =
a^{-n} \frac{\partial}{\partial t} (a^n \dot{\phi}).
\end{equation}
Thus, we introduce an integration constant $C_0 = a^n \dot{\phi}$. The
familiar Friedmann constraint (\ref{e1}) may then be rewritten as
\beq
\labell{FC0}
-k=\dot{a}^2 +w_{eff}(a)\qquad{\rm where}\quad w_{eff}(a)=-\frac{C_0^2}{a^{2(n-1)}}
- V_0 a^2\ .
\eeq
Hence we may view the dynamics of the scale factor $a$ as that of a classical particle
moving in a potential $w_{eff}$ with energy $-k$.

Note that if $C_0\not=0$, the effective potential is
manifestly negative. However,
for the flat and hyperbolic cases ($k=0,-1$), the effective energy is non-negative.
Hence for all such solutions, the scale factor reaches zero in either the past or
future. Further with $C_0\not=0$, one finds a curvature singularity at this point:
$a\sim t^{1/n}$ and $R\simeq -{n\over n+1} t^{-2}$. Hence these solutions correspond
either to a universe which begins in a contracting dS phase
and for which there is enough energy density in the evolving
scalar field to produce a big crunch singularity, or similarly to a universe
which emerges from a big bang to evolve into an expanding dS phase.
Of course, with $C_0=0$, the scale factor $a$ reaches zero in a non-singular
way which simply corresponds to a horizon in dS space, as discussed above.

\FIGURE{\epsfig{file=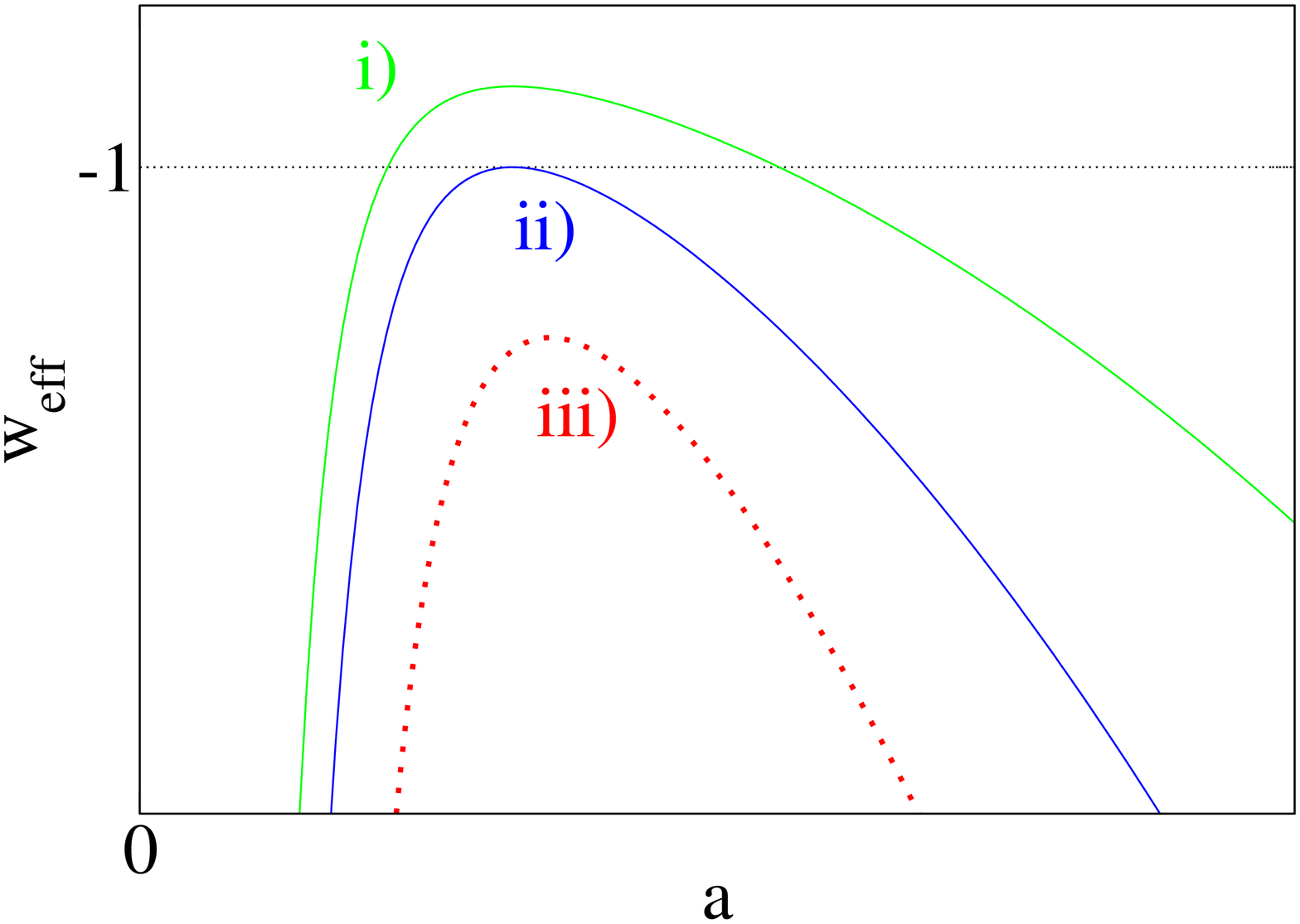, width = 7cm}\caption{The
effective potential $w_{eff}$ with $n=3, k=+1$ for three different cases:
i) small $C_0^2V_0^2$, ii) $C_0^2V_0^2 = \frac{4}{27}$,
and iii) large $C_0^2V_0^2.$ 
} \label{effpot}}

However, the spherical case ($k{=}+1$) is more interesting since the effective
energy is negative. For $C_0^2 > V_0^{-(n-1)} \frac{1}{n-1}
\left(\frac{n-1}{n} \right)^n$, one has $w_{eff}<- 1$ for all values of $a$.
Hence the solutions in this range have a similar interpretation as above
with a big crunch or big bang singularity. With $C_0^2 =
V_0^{-(n-1)} \frac{1}{n-1}\left(\frac{n-1}{n} \right)^n$, the peak of
the effective potential is precisely $-1$, and so there are various classes
of solutions: i) the scale factor is constant with $a^{2n}=V_0^{-n}
\left(\frac{n-1}{n} \right)^n$ and the geometry corresponds to an
Einstein static universe, ii) the universe emerges from a big bang
and asymptotically evolves toward the previous static geometry,
iii) the universe begins with a contracting dS phase and asymptotically
approaches the static phase above, and iv) the time reversal of the
solutions in either (ii) or (iii). For smaller values of $C_0$, the universe is
confined either to small $a$ (leading to solutions having both a
big bang and a big crunch) or to large $a$ (with dS phases both
to the past and future). In all cases, $\phi$
rolls monotonically in one direction.

A piece-wise constant potential can now be dealt with by patching together
solutions described by the above effective potentials.
When $\phi$ crosses a jump in the potential $V$,
the equations of motion show that $a$,
$\dot{a}$, and $\phi$ are continuous but that $\dot{\phi}$ and $\ddot{a}$
suffer a discontinuity.  The Friedmann constraint \reef{e1} then shows that this
discontinuity is determined by the requirement that the scalar
field energy density $\dot{\phi}^2 + V$ is conserved through the transition.
That is, the two solutions are pasted together such that the effective
potential is continuous across the join.  As a result, the
integration constants $C_1,C_2$ associated with potentials
$V_1,V_2$, respectively satisfy the constraint:
\beq
C_2^2+V_2a^{2n} = C_1^2 + V_1 a^{2n}\ .
\labell{cross}
\eeq
This analysis will be used in the construction of a `very tall'
universe in section \ref{construct}.

\section{A generalized de Sitter c-theorem}
\label{ct}

Having illustrated a framework in which we may study
asymptotically dS spaces which are more general than the original
dS spacetime, we would like to consider the role of such solutions
in the context of the dS/CFT correspondence \cite{AS1}. Much of
the development of this duality relies on intuition developed in
studying the AdS/CFT correspondence \cite{adsrev}. One of the
interesting features of the latter is the UV/IR correspondence
\cite{UVIR,UVIR2}. That is, physics at large (small) radii in the AdS
space is dual to local, ultraviolet (nonlocal, infrared) physics
in the dual CFT. As was extensively studied in gauged
supergravity --- see, \eg \cite{flows} --- `domain wall' solutions which evolve
from one phase near the AdS boundary to another in the interior
can be interpreted as renormalization group flows of the CFT when
perturbed by certain operators. In analogy to Zamolodchikov's
results for two-dimensional CFT's \cite{zam}, it was found that a
c-theorem could be established for such flows \cite{ctheo} --- see
also \cite{ctheo2} --- using Einstein's equations. The c-function
defined in terms of the gravity theory then seems to give a local
geometric measure of the number of degrees of freedom relevant for
physics at different energy scales in the dual field theory.

In the dS/CFT duality, there is again a natural correspondence
between UV (IR) physics in the CFT and phenomena occurring near
the boundary (deep in the interior) of dS space. In the context then
of more general solutions which are asymptotically dS, one has an
interpretation in terms of renormalization group flows, which
should naturally be subject to a c-theorem \cite{AS,BBM}.
The original investigations \cite{AS,BBM} considered only solutions
with flat spatial sections ($k=0$), and we generalize these results
in the following to include spherical and hyperbolic sections ($k=\pm1$).
We also consider the flows involving anisotropic scalings
of the boundary geometry, but our results are less conclusive in this case.

\subsection{The c-function}

The foliations of spacetimes of the form given in eq.~\reef{ansatz2c}
are privileged in that time translations a) act as a
scaling on the spatial metric, and thus in the field theory dual and b)
preserve the foliation and merely move one slice to another.
In the context of the dS/CFT correspondence, these
properties naturally lead to the idea that time evolution in
these spaces should be interpreted as a renormalization group flow \cite{AS,BBM}.
Certainly, the same properties apply for time evolution independent of
the curvature of the spatial sections, and in fact also apply
(in the asymptotically dS regions) for any of the metrics presented in
Appendix \ref{folb}. Hence if a c-theorem applies for the $k=0$ solutions
\cite{AS,BBM}, one might expect that it should extend to these other
cases if properly generalized.

For $k=0$, the proposed c-function \cite{AS,BBM}, when generalized to
$n{+}1$ dimensions, is
\beq
\labell{oldc}
c \simeq {1\over\GN\left| {\dot{a}\over a}\right|^{n-1}}\ .
\eeq
The Einstein equations ensure that $\prt_t\left(\dot{a}/a\right)<0$,
provided that any matter in the spacetime satisfies the null energy
condition \cite{HawEll}. This result then guarantees that $c$ will
always decrease in a contracting phase of the evolution or increase
in an expanding phase.

For our general study, we wish to define a c-function which can be evaluated
on each slice of some foliation of the spacetime.
Of course, our function should satisfy a `c-theorem', \eg our
function should monotonically decrease as the surfaces contract in the
spacetime evolution. Further, it should be a geometric function built from the
intrinsic and extrinsic curvatures of a slice. Toward this end, we begin with the
idea that the c-function is known for any slice of de Sitter
space, and note that in this case, eq.~\reef{oldc} takes the form
\begin{equation}
\labell{c}
c \sim {1\over\GN\Lambda^{(n-1)/2}}\ .
\end{equation}
Thus, if our slice can be embedded in some de Sitter space
(as was shown to be the case for any isotropic homogeneous slice
in section \ref{fol}), the
value of the c-function should be given by eq.~(\ref{c}).  In other words, we
can associate an effective cosmological constant $\Lambda_{eff}$
to any slice that can be embedded in de Sitter space and we can
then use this $\Lambda_{eff}$ to define our c-function.

It is useful to think a bit about this embedding in order to
express $\Lambda_{eff}$ directly in terms of the intrinsic and
extrinsic curvatures of our slice. The answer is readily apparent
from the general form of the `vacuum' Einstein equations with
a positive cosmological constant: $G_{ij} = -\Lambda g_{ij}$.
Contracting these equations twice along the unit normal $n^i$ to the
hypersurface gives the Hamiltonian constraint, which is indeed a
function only of the intrinsic and extrinsic curvature of the
slice\footnote{The momentum constraints vanish in a homogeneous
universe, and time derivatives of the extrinsic curvature only appear in the
dynamical equations of motion.}. The effective
cosmological constant defined by such a local matching to de Sitter
space is therefore given by
\begin{equation}
\labell{Leff} \Lambda_{eff} = G_{ij}n^i n^j.
\end{equation}
For metrics of the general form (\ref{metric2b}), this becomes
\begin{equation}
\labell{spec} \Lambda_{eff} = \frac{n(n-1)}{2}\left[
\left(\frac{\dot{a}}{a}\right)^2 +{k\over a^2}\right].
\end{equation}
Taking the c-function to be a function of this effective
cosmological constant, dimensional analysis then fixes it to be
\begin{equation}
\labell{cc}
c \sim \frac{1}{\GN\Lambda_{eff}^{(n-1)/2}} = \frac{1}{\GN}\left({G_{ij}
n^i n^j} \right)^{-(n-1)/2}\ .
\end{equation}
For the $k=0$ isotropic case, it is clear that this reduces to the
c-function \reef{oldc} given previously in \cite{AS,BBM}.  For other isotropic
cases, it is uniquely determined by the answer for the
corresponding slices of de Sitter space.  The same holds for an
anisotropic slice that can be embedded in de Sitter (see e.g. Appendix
A for examples).  While the choice 
(\ref{cc}) is not uniquely determined by the
constraints imposed thus far for any slice which cannot be so embedded,
it does represent a
natural generalization and, as we will see below, this
definition allows a reasonable `c-theorem' to be proven.

\subsection{The c-theorem}

For any of the homogeneous flows as considered in the previous section, it
is straightforward to show that our c-function \reef{cc} always decreases
(increases) in a contracting (expanding) phase of the evolution. However,
we would like to give a more general discussion which in particular
allows us to consider
anisotropic geometries, as well as these isotropic cases.

To prove our theorem, we note that the Einstein equations relate
our effective cosmological constant to the energy density $\rho$
on the hypersurface,
\begin{equation}
\Lambda_{eff} =G_{ij}\,n^i n^j =
T_{ij}\,n^i n^j = \rho.
\end{equation}
Consider now the `matter energy' $E = \rho V$ contained in the
volume $V$ of a small co-moving rectangular region on the
homogeneous slice. That is, we take
\begin{equation}
V =\int_R \sqrt{g}\, d^n x
\end{equation}
for some small co-moving region $R$ of the form $R = \{ x | x^i_a
<x^i < x^i_b \}$ where $x^i$ denote co-moving spatial coordinates.
We also introduce $\delta x^i = x^i_b-x^i_a,$ the co-moving size
of $R$ in the $i$th direction.   Since $R$ is small, each
coordinate $x^i$ can be associated with a scale factor $a^i(t)$
such that the corresponding physical linear size of $R$ is
$a^i(t)\delta x^i.$

Without loss of generality, let us assume that the coordinates
$x^i$ are aligned with the principle pressures $P_i$, which are
the eigenvalues of the stress tensor on the hypersurface.   Let us
also introduce the corresponding area $A_i$ of each face. Note
that a net flow of energy into $R$ from the neighboring region is
forbidden by homogeneity.  As a result, energy conservation
implies that $dE = - P_i A_i
d(a\delta x^i)$ as the slice evolves. However, clearly $dE = \rho dV +
V d \rho$,  so that we have
\begin{equation}
\labell{dL} d\Lambda_{eff} = d \rho = -\sum_i \left( \rho +
P_i\right) d\ln a_i.
\end{equation}
Now we will assume that any matter fields satisfy the weak energy
condition so that $\rho~+~P_i~\ge~0$.  Thus, if all of the scale
factors are increasing, we find that the effective cosmological
constant can only decrease in time.

This result provides a direct
generalization of the results of \cite{AS,BBM} to slicings that
are not spatially flat.  In particular, in the isotropic case
(where all scale factors are equal, $a \equiv  a_i = a_j$), it
follows that $c(a)$ as given in eq.~\reef{spec} is, as desired, a monotonically
increasing function in any expanding phase of the universe.

Note, however, that the anisotropic case is not so simple to interpret. For
example, it maybe that the scale factors are expanding in some
directions and contracting in others.   In this case our effective
cosmological constant may either increase or decrease, depending
on the details of the solution.

\subsection{Complete Flows versus Bouncing Universes}

The general flows are further complicated by the fact that they may
`bounce', \ie the evolution of the scale factor(s) may reverse itself.
The simplest example of this would be the $k{=}+1$ foliation of dS space
in section \ref{fol}. In this global coordinate system, the scale factor
\reef{radd} begins contracting from $a(t=-\infty)=\infty$ to $a(t=0)=1$,
but then expands again toward the asymptotic region at $t=+\infty$. In contrast,
we refer to the $k{=}0$ and --1 foliations as `complete'.
By this we mean that within a given coordinate patch, the flow proceeds
monotonically from $a = \infty$ in the asymptotic region to $a=0$ at the boundary
of the patch --- the latter may be either simply a horizon (as in the case of
pure dS space) or a true curvature singularity.

For any homogeneous flows, such as those considered in section \ref{gass}, it is not
hard to show that the $k{=}0$ and --1 flows are always complete and that only the
$k{=}+1$ flows can bounce. The essential observation is that for $a(t)$ to bounce
the Hubble parameter $\dot{a}/a$ must pass through zero. Now the ($tt$)-component
of the Einstein equations \reef{eomf} yields
\beq
\labell{ttcom}
\left({\dot{a}\over a}\right)^2=T_{tt}-{k\over a^2}\ .
\eeq
Now as long as the weak energy condition applies,\footnote{Note
that if $k=0$ and the energy density is identically zero, it follows that
$a$ is a constant. Hence in this case, we will not have an asymptotically
dS geometry.} it is clear that the right-hand-side is always positive for
$k=0$ and --1 and so $\dot{a}/a$ will never reach zero. On the
other hand, no such statement can be made for $k{=}+1$ and so it is only
in this case that bounces are possible. Further one might observe that this
analysis does not limit the number of bounces which such a solution might
undergo. In certain cases with a simple matter content, \eg dust
or radiation, one may show that only a single bounce is possible. However in
(slightly) more complex models, multiple bounces are possible. We will
illustrate this behavior in section \ref{construct}, where a solution
with a rolling scalar is constructed with multiple bounces --- see also
appendix \ref{bouncers}.
Finally in the case of anisotropic solutions --- see, \eg Appendix \ref{folb}
--- the characterization of the flows as complete or otherwise is more
complicated.

The renormalization group interpretation
of bouncing universes is certainly less straightforward. Perhaps greater
insight into this question can come from further study of renormalization group
flows and the UV/IR correspondence in the AdS/CFT context for
foliations of AdS space where the sections have negative curvature.
Such AdS solutions show a similar bounce behavior
--- see, \eg \cite{us}.

\section{The global perspective}
\label{diagrams}

Recall from the introduction the observation that
asymptotically de Sitter conformal diagrams are `tall', \ie
an entire compact Cauchy surface will be visible to
observers at some finite time, and hence that
perturbations of dS space may bring features that originally lay behind a
horizon into an experimentally accessible region. Specifically,
these results rely on Corollary 1 of \cite{GW}, which we paraphrase as follows:

\begin{quotation}
Let the spacetime $(M,g_{ij})$ be null geodesically complete and satisfy
the weak null energy condition and the null generic condition.  Suppose
in addition that $(M,g_{ij})$ is globally hyperbolic with a compact
Cauchy surface $\Sigma$.  Then there exist Cauchy surfaces $\Sigma_1$
and $\Sigma_2$ (of the same compact topology, and with $\Sigma_2$ in the
future of $\Sigma_1$) such that if a point $q$ lies in the future
of $\Sigma_2$, then the entire Cauchy surface $\Sigma_1$ lies in the
causal past of $q$.
\end{quotation}

\noindent
This is sufficient to guarantee that the conformal diagram is `tall'
in the sense of figure \ref{heights} (b).  While it need not necessarily be
`very tall' in the sense of figure \ref{heights} (c), the possibility
is open that this may hold in certain cases so that
a compact Cauchy surface may actually
be found to lie within the
intersection of the past {\it and} future of a generic worldline.
An example of such a spacetime will be constructed in section \ref{construct}
below.

The diagrams in figure \ref{heights} were not intended to
represent  generic asymptotically de Sitter spacetimes.  Instead, these
diagrams only illustrate what is meant by `tall' and `very tall' spacetimes.
The purpose of the current section is to
construct the general such conformal diagrams corresponding to our homogeneous
flows.  This may be of particular interest if in the end
there is a meaningful dS/CFT correspondence in which (either
distinct or isomorphic) field theories are associated with both
$I^+$ and $I^-$.  We note that two field theories are indeed of
relevance to certain applications \cite{LM,Juan} of the more
developed AdS/CFT correspondence.

Although we have shown that the evolution of $c(a)$ is monotonic, certain unusual
features of our flow become apparent when we study the global
structure of the spacetimes dual to our field theory. Let us assume
that the slices are isotropic and take each of the three possible
cases (spheres, flat slices, and hyperbolic slices) in turn.

\subsection{Flat slices ($k=0$)} \label{flatslice}

We now wish to construct the conformal diagram for flows with
flat spatial sections. In order to draw useful two-dimensional
diagrams, we shall use the common trick of studying rotationally
symmetric spacetimes and drawing conformal diagrams associated
with the `$r$-$t$ plane,' \ie associated with a hypersurface
orthogonal to the spheres of symmetry.

\FIGURE{\epsfig{file=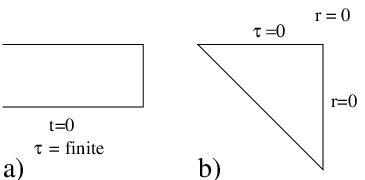}\caption{In the distant past $\tau$ is a)
finite b) infinite. }\label{flat}}

For later use, we begin with an arbitrary $n+1$ dimensional
spatially homogeneous and spherically symmetric metric in proper
time gauge:
\begin{equation}
ds^2 = - dt^2 + a^2(t) \left( dr^2 + \hat R^2(r) d \Omega^2_{n-1}
\right),
\end{equation}
where $d\Omega^2_{n-1}$ is the metric on the unit $n-1$ sphere and
the form of $\hat R(r)$ depends on the spatial geometry: $\hat
R(r) = \sin(r), r, \sinh(r)$ for spherical, flat, and hyperbolic
geometries respectively.  In fact, the function $\hat R(r)$ will
not play a role below as our diagrams will depict the conformal
structure only of the ($1+1$)-dimensional metric $ds^2_{1+1} = - dt^2 + a^2(t) dr^2.$
However, it will be important to note that $r$ takes values only
in $[0,\pi]$ for the spherical geometry but takes values in
$[0,\infty]$ for the flat and hyperbolic cases.
The usual change of coordinates to conformal time $\tau(t)$ defined by $d\tau =
\frac{dt}{a}$ leads to the conformally Minkowski metric
\begin{equation}
\labell{cgeom} ds^2_{1+1} = a^2(t)(-d\tau^2 + dr^2).
\end{equation}

Let us assume that our foliation represents an expanding phase
that is asymptotically de Sitter in the far future.  That is, for
$t \rightarrow + \infty$ the scale factor $a$ diverges
exponentially.
There are now two possibilities.  Suppose first that $a =0$ at
some finite $t$.   If $a^{-1}$ diverges as a small enough power of $t$ then
$\tau$ will only reach a finite value in the past and the
spacetime is conformal to a half-strip in Minkowski space.

In contrast, if $a$ vanishes more quickly or if it vanishes only
asymptotically then $\tau$ can be chosen to take values in
$[-\infty,0]$.  From (\ref{cgeom}) we see that the region covered by
our foliation is then conformal to a quadrant of Minkowski space.  We
take this quadrant to be the lower left one so that we may draw
the conformal diagram as in figure \ref{flat} (b).

We now wish to ask whether the region shown in figure \ref{flat} (b)
is `complete' in
some physical sense.  In particular, we may wish to know whether
light rays can reach the null `boundary' in finite
affine parameter.  A short calculation shows that the affine
parameter $\lambda$ of a radial null ray is related to the original
time coordinate $t$ by $d \lambda = a dt.$  The affine parameter
is clearly finite if $a$ vanishes at finite $t$.
In the remaining case, we have seen
that $\rho$ is bounded below. As a result, $a$ must vanish at
least exponentially and the affine parameter is again finite.

\FIGURE{\epsfig{file=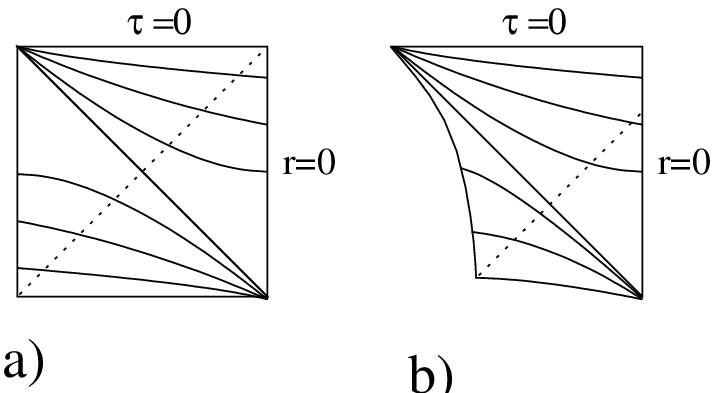, width=8cm}\caption{a) In a square diagram, a light
ray reaches the antipodal point only at $I^+$.  b) The
generic conformal diagram for an asymptotically
de Sitter space with flat surfaces of homogeneity.  A light ray
starting in the lower left corner reaches the antipodal point
at a finite time.}\label{genflat}}

Thus, this null surface represents either a singularity or
a Cauchy horizon across which our
spacetime should be continued. 
This statement is essentially a restricted version of the results of
\cite{BV1,BV2,BV3,BGV} (see also \cite{VT} for other interesting
constraints on the `beginning' of inflation).  
Note that there is no tension between
our possible Cauchy horizon and the claim of a ``singularity''
in these references, as their use of the term singularity
refers only to the geodesic incompleteness of the expanding phase.

From (\ref{dL}) we see that unless $\rho+P_i$ vanishes as $a \rightarrow 0$, 
the energy density must diverge and a curvature singularity will indeed
result.  However, a proper tuning of the matter fields 
can achieve a finite $\rho$ at $a=0$.  It is therefore interesting to  
consider solutions which are
asymptotically de Sitter near $t=0$ so that $a$
vanishes exponentially.  In this case, the $a=0$ surface represents a Cauchy
horizon across which we should continue our spacetime.
We will focus exclusively on such cases below.

Since the boundary is a Cauchy horizon, there is clearly some
arbitrariness in the choice of extension. We make the natural assumption
here that the spacetime beyond the horizon is again foliated
by flat hypersurfaces.  Although at
least one null hypersurface ($t=-\infty$) will be required,
it can be shown that the surfaces of homogeneity must again become
spacelike across the horizon if the spacetime is smooth.
The key point here is
that the signature can be deduced from the behavior of $a^2(t)$,
which gives the norm $|\xi|^2$ of any Killing vector field $\xi$
associated with the homogeneity.  We impose a ``past asymptotic de
Sitter boundary condition'' so that the behavior of this quantity
near the Cauchy horizon must match that of some de Sitter
spacetime.  Consider in particular the behavior along some null
geodesic crossing the Cauchy horizon and having affine parameter
$\lambda$.  It is straightforward to verify that $\lambda \sim a$,
so that matching derivatives of $|\xi|^2$ across the horizon
requires $\xi$ to again become spacelike beyond the horizon.

It follows that the region beyond the Cauchy surface is just
another region of flat spatial slices, but this time in the
contracting phase. It is therefore conformal to the upper right
quadrant of Minkowski space. However, having drawn the above
diagram for our first region we have already used a certain amount
of the available conformal freedom. Thus, it may not be the case
that the region beyond the Cauchy surface can be drawn as an
isosceles right triangle.  The special case where this is possible
is shown in figure \ref{genflat} (a).  The exceptional nature
of this case can be seen from the fact that it allows
a spherical congruence of null geodesics to proceed
from the upper right corner of $I^+$ (where it would have zero
expansion) to the lower right corner of $I^-$ (where it would also
have zero expansion).  Assuming as usual the weak energy
condition, it follows that this congruence encountered no focusing
anywhere along its path; i.e., $\rho + P =0$. Given the high
degree of symmetry that we have already assumed, this can happen
only in pure de Sitter space. The correct diagram for the general
case is shown in figure \ref{genflat} (b) (see appendix
\ref{flatdiagram} for a complete derivation).

\FIGURE{\epsfig{file=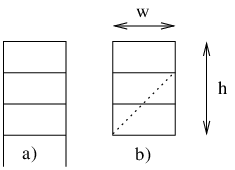,width=5cm}\caption{Conformal
diagrams for spherical surfaces
of homogeneity a) for the case where $\tau$ diverges in the past and
b) for the case where $\tau$ converges in the past.}\label{ss}}

\subsection{Spherical slices ($k=+1$)}

The conformal diagrams in this case are relatively simple. Since the
radial coordinate now takes values only in an interval, we see
from (\ref{cgeom}) that the conformal diagram is either a rectangle
or a half-vertical strip, depending on whether or not $\tau$ is
finite at the past boundary.

All such rectangles with the same ratio h/w (see figure
\ref{ss} (b)) can be mapped into
each other via conformal transformations.  For the case of pure de
Sitter space we have h=w.  On the other hand, for any spacetime
satisfying the generic condition (so that null geodesics suffer
some convergence along their trajectory), we know from
\cite{GW} that the region to the past of any point
$p$ sufficiently close to $I^+$ must contain an entire Cauchy
surface. Thus\footnote{This conclusion may also be reached by
considering the sphere of null geodesics that begins in, say,
the lower left corner and progresses toward the upper right and using the
non-increase of the expansion $\theta$ implied by the weak null energy
condition.}, for such cases we have h $>$ w. Similar `tall' spacetimes
were recently considered in the context of cosmologies violating the
dominant energy condition \cite{brett}.

\subsection{Hyperbolic slices ($k=-1$)}
\label{hyp}

\FIGURE{\epsfig{file=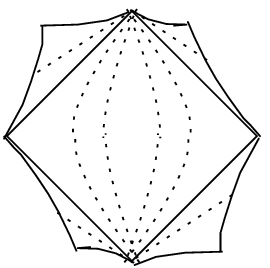, width=4cm}\caption{The general conformal
diagram for an appropriately
complete asymptotically de Sitter spacetime with hyperbolic surfaces
of homogeneity.  A conformal frame has been chosen
such that the diagram has a $Z_2$ reflection symmetry through the center.}
\label{hyp2}}

Recall that the hyperbolic flows are complete, \ie  $a$ reaches
0 at finite $t$ (say, $t=0$), and vanish at least as fast as $t$.
The asymptotically de Sitter boundary conditions also
require that $a$ diverge exponentially as $t \rightarrow
+\infty$. Note that since $a$ vanishes quickly, $\tau$ will
diverge at $t=0$ and the region is again conformal to a quadrant
of Minkowski space. As usual, this may or may not be singular
depending on the matter present.

Consider in particular the asymptotically de Sitter case where $a$
vanishes linearly. One then finds that the affine parameter
$\lambda$ along a null ray near the horizon is asymptotically
$\lambda \sim a^2$.  The Killing vector field that implements spatial
translations in the direction along this null ray has a norm given
by $a^2 \sim \lambda$ along this null
ray and so, if the spacetime is smooth,
must become timelike beyond the horizon. Thus, the homogeneous
surfaces must become timelike on the other side. As can be seen
from (\ref{tlh}), an asymptotically de Sitter region foliated by
timelike hyperbolic slices (i.e., copies of de Sitter space) has
an `r-t plane' that is conformal to a diamond in Minkowski space.

Assuming that no singularities are encountered within this diamond or
on its boundaries, this
provides three further Cauchy horizons across which we would
like to extend our spacetime.  A study of the norm of the Killing
fields tells us that the foliation must again become spacelike
beyond these horizons.  Just as we saw for the flat foliations, we are
therefore left with the task of attaching pieces conformal to
various quadrants of Minkowski space. By the same reasoning as in
appendix \ref{flatdiagram}, the complete conformal diagram can be
drawn as shown in figure \ref{hyp2}.

\section{`Very tall' universes}
\label{construct}

This section is devoted to constructing an asymptotically de
Sitter spacetime whose conformal diagram is `tall' enough that a
compact Cauchy surface can be experimentally probed by a co-moving
observer.  In particular, such a Cauchy surface will lie in the
intersection of the observer's past and future. We refer to such
spacetimes as `very tall.'  The solutions we
construct below
will all satisfy the dominant energy condition. Interestingly, we
will discover spacetimes of this form for which the spatial volume
of the visible Cauchy surface can be made arbitrarily large.

\FIGURE{\epsfig{file=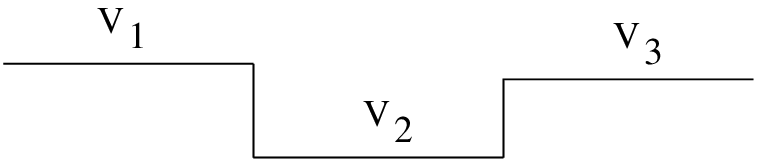}
\caption{Our piece-wise constant potential.}\label{ttuPOT}}
Our solution will again be built by considering
a scalar field $\phi$ interacting
with Einstein-Hilbert gravity as described by the action \reef{totact}.
We will be using the spatially homogeneous ansatz \reef{ansatz2c}
with spherical sections ($k{=}+1$). The scalar potential is
taken to be non-negative and
piece-wise constant, and so the dynamics of this system are
of the form discussed in section \ref{step}. In particular,
we consider a potential of the form shown in figure \ref{ttuPOT}, which
consists of three steps of heights $V_{1,2,3}$.
We will be interested in the case where $V_1$ and $V_3$ are comparable,
though not necessarily equal, and $V_2<V_{1,3}$. There will be
three corresponding constants of integration $C_{1,2,3}$
for eq.~(\ref{phieq0}) that will be related by the matching condition
\reef{cross}. Given these constants, the evolution of the scale factor
on each of the three steps corresponds to that of a classical particle
with energy $-1$ in the corresponding effective potential, $w_{1,2,3}$,
as given in eq.~\reef{FC0}.

Suppose that we choose $C_1^2 < V_1^{-(n-1)} \frac{1}{n-1}
\left(\frac{(n-1)}{n} \right)^n$ so that the effective
potential satisfies $w_1>-1$ for some values of the scale
factor.  If we begin the universe
in an asymptotically de Sitter contracting phase,
the universe will reach some locally minimum
size $a_1$ and then bounce off of the potential. That is, $a$ is
confined to values {\it larger} than those where $w_{1}\le-1$.
Suppose also that we arrange the initial conditions so
that $\phi$ rolls from $V_1$ down to $V_2$ when the universe is at
some size $a_{12} > a_1.$ We will imagine keeping $V_1$, $C_1$,
and $a_{12}$ fixed and then tuning $V_2$ to attain the desired
behavior in the following.

Now the matching condition \reef{cross} yields
$C_2^2 = C_1^2 + (V_1 -V_2)a_{12}^{2n} > C_1^2$.
It is clear that by taking $V_2$ sufficiently small we will have
$C_2^2 < V_2^{-(n-1)} \frac{1}{n-1}\left(\frac{n-1}{n} \right)^n$ so that
the effective potential rises above $-1$. Furthermore, a study of
(\ref{FC0}) shows that $w_2$ rises above the value -1 only
at some $a_2>a_{12}$.  Hence on the second step, the solution
will continue expanding for a while, but eventually it bounces off
the new effective potential to enter a contracting phase.
Physically, the scalar field has attained so much kinetic energy
by rolling from $V_1$ down to $V_2$ that this kinetic energy now dominates
the dynamics and the spacetime will expand to a locally maximum
size $a_2$ and then begin to re-collapse again, much as in a
standard spherical matter-dominated FRW cosmology.

To reach a final expanding asymptotically de Sitter phase, we need
only run the construction in reverse.  In particular, we need only
arrange the scalar field to roll up to $V_3$ when the scale factor
takes some value $a_{23}$ close to $a_{12}$ and to take $V_3$
close to $V_1$.  In this case the cosmological constant again
takes over, creating another bounce at some locally minimum size
$a_3$.  After this point, the spacetime expands forever and
asymptotically approaches a de Sitter solution.

It is also clear that one could in principle alternate de Sitter
and FRW-like phases indefinitely. Note that by taking $a_{23}$
{\it smaller} than $a_{12}$ the same behavior can be obtained by
choosing a $V_3$ that is larger than $V_1$.   In this way, one can
construct a spacetime that is asymptotically de Sitter to the past
and future with arbitrary independent cosmological constants
$\Lambda_{initial}$ and $\Lambda_{final}$ and which nevertheless
oscillates arbitrarily between large and small sizes.

While some amount of tuning is required in our construction, it is clear
that the scenario is stable in the sense that there is an open set of
parameter space that leads to the desired behavior. For the reader
who may find our analysis with a piece-wise scalar potential lacking,
we also present some numerical calculations illustrating analogous
bouncing universes for a smooth potential in Appendix \ref{bouncers}.
It should also be
clear that we may choose the size $a_2$ and the duration of the
internal FRW region as large as we wish. Hence these `very tall' universes
may also be `big wide' universes.

\subsection{Entropy in `big wide' universes}
\label{bwentropy}

The `big wide' universes constructed above are particularly
interesting in the light of recent discussions
in which the fact that the cosmological horizon prohibits an observer in de
Sitter space from accessing more than a fixed, finite spatial
volume was used to motivate the idea that asymptotically de
Sitter spaces contain only a finite number of degrees of freedom \cite{Banks}.
Given such arguments, the construction of big wide
asymptotically dS universes satisfying the weak and
dominant energy conditions may come as a surprise.
Certainly, any standard field theory given access to a fixed
energy and an arbitrarily large volume will exhibit an arbitrarily
large number of degrees of freedom.

Scenarios of this kind may certainly be envisioned in a big wide
universe and they are instructive to explore.  Suppose for example
that we add a single massive particle of small but fixed mass to
such a spacetime, where the mass $m$ is chosen small enough that
the particle has negligible effect on the gravitational dynamics
through the first de Sitter bounce (at $a_{12}$) and into the
middle expanding phase.  We may imagine that this particle is
coupled to some massless field (say, another scalar field), into
which it may decay when the universe is very large.  This in
principle supplies the scalar field with a fixed
amount $m$ of energy in a large volume.  In some sense then this
system does `have' an arbitrarily large number of available
states.

Let us investigate, however, what happens when those states are
actually accessed.  The particle decays and some small amount of
massless radiation is excited in a universe much larger than the
length scale $l_{final}$ set by the final cosmological constant.
Our big wide universe will eventually reach the maximum size
($a_2$) of its FRW-like expansion and begin to recollapse.  While
the expansion or contraction of the spacetime had little effect on
the massive progenitor particle, the contraction will significantly
blueshift massless radiation in proportion to the inverse scale
factor ($1/a$).  For example, one might approximate this massless radiation
by a thermal state, in which case the temperature will increase as the spacetime
contracts.  As a result, if this entropy survives to reach
the second de Sitter bounce at $a_{12}$ the matter energy density
is now much greater than it was during the first bounce (roughly
by a factor of $a_2/a_{12})$.  It may therefore be large enough to
dominate the gravitational dynamics and drive the universe into a
big crunch instead of a second re-expansion.

That a big crunch would indeed be the outcome can be said with
some confidence. Let us recall that Bousso's `covariant entropy
bound' \cite{covBound} can be proven to hold \cite{FMW} in the
context of classical general relativity coupled to matter
satisfying the restrictions
\begin{equation}
\labell{FMWcond} T_{ij}k^ik^j \ge c_1 \ell_p^{-1} |k_i s^i|^2, \
\ \ \ \ \ \ \ \ \ \ \ T_{ij}k^ik^j \ge c_2 k^i k^j \nabla_i s_j,
\end{equation}
where $k^i$ is an arbitrary null vector and $T_{ij}$ and $s^i$ are
the matter stress-energy tensor and entropy current,\footnote{While
the fundamental status of such a notion may be unclear, we will
see that it is sufficiently well-defined in the current context.}
respectively.  The constants $c_1$ and $c_2$ are coefficients of
order $1$ and we have explicitly indicated factors of the Planck
length $\ell_p$. 
The theorem is then that the integral $\oint
k_is^i\, dS$ over any non-expanding null surface $S$ is bounded by
$A/4\ell_p^{n-1}$, where $A$ is the area of its initial cross-section.
It was argued in \cite{FMW} that subject to a restriction
on the number of
species, (\ref{FMWcond}) holds for any matter below the Planck
temperature whose entropy can be successfully modeled by a
classical entropy current of the sort used in fluid dynamics.
As noted above, it suffices
to consider the
evolution of the universe that would result if our scalar field is place
in a thermal state at the moment the massive particle decays.

Suppose that the entropy of this thermal state is more
than\footnote{The unusual factor of two is not important here. It
may be of interest for other purposes, but most likely it is a
product of our inefficient estimate below of the back reaction of
the thermally excited field on the metric. In any case, adjusting
the constants $c_1$, $c_2$ in (\ref{FMWcond}), and thus
restricting the temperature to be a factor of order one below the
Planck scale, can easily be made to insert factors of 2 into the
bound on the entropy flux.} $2A/4\ell_p^{n-1}$, where $A$ is the area
of the de Sitter horizon associated with the final cosmological
constant $\Lambda_{final}$ in the original big wide universe. For
the discussion below we will assume that all temperatures remain
below the Planck scale so that we can make predictions with some
confidence.

The recollapse of the universe acts on the massless field just as
if we studied a field inside a box that is being compressed.  The
compression is adiabatic as the gravitational field does not
exchange heat with the thermally excited scalar field.  As a
result, the entropy of each co-moving volume element will remain
constant.

Supposing that the universe has an asymptotically de Sitter final
phase with the same final effective
cosmological constant $\Lambda_{final}$ as in
our original (tall!) big wide universe will now lead to a
contradiction. Consider the final bounce and in particular the
spherical light front emitted from the equator of the sphere at
this time.  There are two such light fronts, one emitted toward
either pole.  Let us arbitrarily choose the one emitted toward the
north pole.

Such a null congruence necessarily begins with non-positive
expansion since $\dot{a}$ vanishes on the slice and, within the given
slice, there is no larger surface toward which to expand. The weak energy
condition then guarantees that the congruence must continue to
contract until it reaches the north pole.  Thus, this null surface
satisfies the conditions for the application of the covariant
entropy bound \cite{covBound}.

Note that $\dot{a}$ vanishes at the final bounce and that the
matter energy density is greater than in the case where the massive
particle did not decay.
As a result, the Friedmann constraint  \reef{e1} guarantees that
the minimal sphere is {\it smaller} and the initial area of our
light sheet is less than $A/4\ell_p^{n-1}.$ However, we see that half
of the matter entropy must flow through this null surface,
contradicting the theorem of \cite{FMW}. Note that it does not
help to somehow arrange to heat only `the other half of the
universe' as we may simply choose the light front that contracts
toward the south pole.

We see that an attempt to access an entropy larger than that
commonly associated with the final de Sitter phase creates such a
large perturbation that the final de Sitter phase is destroyed.
If one requires the spacetime to have a final asymptotically de
Sitter (expanding) phase with cosmological constant
$\Lambda_{final}$, one finds that the observer is not in fact
allowed to excite more than of order $A_{final}/4\ell_p^{n-1}$ degrees
of freedom, where $A_{final}$ is the area of a de Sitter
cosmological horizon associated $\Lambda_{final}$.

At first sight, this focus on $\Lambda_{final}$ might appear to
lead to a new time asymmetry.  However, this is really just the
familiar thermodynamic arrow of time as we have allowed the
entropy of the matter fields to increase (say, by decay of a
massive particle into massless radiation) but not to decrease with
time.  Of course, we expect that number of degrees of freedom (or
even the entropy ${\rm tr}(\rho \ln \rho)$ of a mixed quantum
state) do not really change at all with the passage of time.
Instead, the increase in entropy is an artifact of choosing a
course-graining of the system in which the initial entropy appears
small.  Thus, barring unexpectedly large violations of unitary
from possible quantum gravity effects, 
the time reverse of our argument suggests that one cannot
prepare the FRW region in more than of order
$A_{initial}/4\ell_p^{n-1}$ states if the spacetime is to pass through
an initial (contracting) asymptotically de Sitter phase with
cosmological constant $\Lambda_{initial}$. The number of
states that can be accessed subject to both boundary conditions is
therefore determined by the larger of $\Lambda_{initial}$ and
$\Lambda_{final}$.

We believe that the terminology used here of
`accessible' versus `available' degrees of freedom or entropy is an
appropriate one. However, one may also take seriously the idea \cite{Banks}
that the fundamental Hilbert space associated to asymptotically de
Sitter space should contain only those states that can in fact be
accessed without destroying the asymptotically de Sitter behavior.

\section{Discussion}
\label{disc}

Our calculations focused on two aspects of asymptotically dS spacetimes
interpreted as renormalization group flows in the dS/CFT correspondence.
One aspect was establishing a generalized c-theorem for these flows.
The other was the construction of conformal diagrams for various
homogeneous flows.

For isotropic flows, we showed that $\Lambda_{eff}$ defines a
c-function (\ref{c}) that is a locally increasing function of
the scale factor $a$ regardless of whether the foliation was by
flat planes, spheres, or hyperbolic spaces. A similar result
applies at least for certain anisotropic cases. In parallel with
the results in the AdS/CFT duality, the present c-theorem essentially
says that the effective cosmological constant is larger in the interior of
the space than at the (conformal) boundary.

The flows associated with flat
and hyperbolic spatial sections were seen to differ markedly
from those on spheres.  In particular, the $k{=}0$ and --1 flows are
always {\it complete}, running monotonically between $a=0$
and $\infty$. In contrast, we showed that the $k{=}+1$ flows
can yield bouncing universes, in which the evolution of the scale
factor may reverse its direction, and in particular produce two
conformal boundaries. In the latter case, while the c-theorem still maintains
that $\Lambda_{eff}$ is larger in the interior of the space than
at the boundaries, it does not establish any relationship between
the $\Lambda_{initial}$ and $\Lambda_{final}$ at $I^-$ and $I^+$,
respectively. In fact by analyzing simple models, it is clear that there
can be no simple relation. For example, considering a single step potential
as discussed in section \ref{step}, we find that for 
a fixed $\Lambda_{initial}$, the 
initial conditions of the scalar can always be chosen such that the universe
succeeds in evolving to an asymptotically dS region as $t\rightarrow+\infty$
for an arbitrarily small $\Lambda_{final}$.

One feature of the $k=0$ foliation which makes the
renormalization group interpretation manifest is the existence of
the Killing vector $\prt_t-{x^i\over\lc}\prt_i$
in pure dS space \cite{AS1}. While this Killing vector naturally acts to
evolve from one time slice to another, it also acts as a global
scale transformation on each slice, and in particular on the
boundary $I^+$. The fact that this flow is a symmetry of
dS space is in keeping with the conformal symmetry
of the dual field theory. While the symmetry is lost for
general $k{=}0$ flows, it is still natural to associate
the time evolution with a flow in energy scales in the dual picture.
For the $k{=}\pm1$ coordinates on de Sitter space, there is
no analogous Killing vector which preserves the foliation of
the spacetime. However, the scaling of the spatial slices that
arises in the time evolution is still suggestive of a renormalization
group flow for such solutions.

As mentioned, the flows with spherical slices
often reach a regime where $\dot{a}$ vanishes
even though $a$ remains finite.  Beyond that point, $\dot{a}$
typically changes sign so that the derivative of our c-function
does as well. That the renormalization group flow cannot go
beyond a finite point may not be a surprise from the point of
view of field theory on the sphere.  In contrast to the flat or
hyperbolic plane, any (simply connected) compact space has a longest length
scale\footnote{In the non-simply connected case, it is well known
that Wilson lines may effectively expand the compact directions
--- see, \eg \cite{bigger,bigger2}. We expect this would play
a role in $k{=}-1$ flows where the hyperbolic slices were compactified
with appropriate identifications.}  and
that, beyond that scale, no useful local effective description of
the theory can be obtained. The surprise is perhaps that the flow
does not just stop, but in fact continues with the c-function
reversing the direction of its course.  In the case where
$\dot{a}$ vanishes at only one sphere of minimal size, it is
natural to interpret the full spacetime as consisting of {\it two}
renormalization group flows, each starting at $I^{\pm}$ and
flowing to the same effective theory at the sphere where $\dot{a}$ vanishes.
It is similarly possible that more complicated spacetimes which
oscillate several times illustrate various complicated
combinations of flows upward and downward in length scales.

Much the same interpretation can be made of the geodesically
complete spacetimes discussed in section \ref{flatslice}, in connection with
$k{=}0$ flows.  If the flow does not proceed to a
singularity at $a=0$, it can be patched across the Cauchy horizon to a
second flow which produces the same result in the IR, \ie produces
the same geometry at the Cauchy horizon. We then see two flows coming
from $I^\pm$ and ending at the Cauchy horizon.

For the flows on hyperbolic spatial slices, we saw that it was
impossible to patch one such flow directly to another across the
Cauchy horizon. Instead, an intermediate region is required in
which the surfaces of homogeneity are timelike, instead of
spacelike, as in eq.~\reef{tlh} for example.
The interpretation of this matching remains obscure to us
and deserves further investigation.

Although anisotropic flows are more complicated to interpret,
much of the above discussion of the isotropic case admits a clear
extrapolation.  If one is willing to sacrifice rotational
invariance then one may course-grain a field theory differently
in different directions. One may use this idea to
construct anisotropic renormalization group flows.  As in the
isotropic case, we regard any surface on which $\dot{a}_i$
vanishes for one of the scale factors as representing the joining
of two flows at some particular scale.  The idea of a flow in
which we move upward in scale in some directions but downward in
scale in the others (i.e., in which the $\dot{a}_i$ do not all
have the same sign) may be unfamiliar, but is certainly allowed if
we adopt the viewpoint advocated above that the coarse-graining
described by our flows in fact keeps track of all of the
information in the more fine-grained theory but that the
c-function describes an effective theory at the scale set by the
$a_i$.

Above we considered joining flows by smoothly matching various
asymptotically dS geometries at a Cauchy horizon. Recall from the
discussion in section \ref{fol} that such
horizons are naturally associated with a(n effective) boundary of the
manifold on which the dual CFT is formulated. Such a boundary appears
because a singular conformal transformation
has been used to push off to infinity
various points in the $S^n$ naturally appearing at $I^\pm$.
In smoothly matching geometries across the horizon, we are implicitly
making a very precise selection for the CFT conditions or geometric
data at these boundaries. This choice, of course, in not unique. For example,
it would be straightforward to match geometries at a Cauchy horizon
so that the derivatives of the metric where discontinuous even though
the metric itself was continuous.  One is also free to break homogeneity
beyond a Cauchy horizon.

A related discussion arises for the singularity conjecture of ref.~\cite{BBM}.
It was pointed out in ref.~\cite{robb} that the negative mass Schwarzschild-dS
solution seems to evade this conjecture in that, while the mass as defined
by \cite{BBM} is greater than that of dS space, there is no `cosmological
singularity.' Rather observers may proceed from $I^-$ to $I^+$ without ever
encountering a singularity in this spacetime. On the other hand, if we consider
situation in terms of evolving Einstein's equations forward from $I^-$, the
maximally analytically continuation of the negative mass Schwarzschild-dS solution
is certainly a very special solution requiring very precise boundary conditions at
$t=\pm\infty$ on $I^-$ (as well as all along the timelike singularity at $r=0$
beyond the Cauchy horizon). Clearly for generic boundary conditions, there will
be discontinuities, \ie impulsive gravitational waves, traveling along the
Cauchy horizon. Hence in the generic situation, observers should be expected
to encounter a `cosmological singularity.' Therefore it seems that the conjecture
of \cite{BBM} may still be valid if refined to include some sort of generic
condition -- or perhaps even just the existence of a single smooth Cauchy
surface.

The c-theorem suggests that the effective number of degrees of freedom in the CFT
increases in a generic solution as it evolves toward an asymptotically
dS regime in the future. We would like to point out, however, that this
does not necessarily correspond to the number degrees of freedom accessible
to observers in experiments. Here we are thinking in terms of holography
and Bousso's entropy bounds \cite{covBound}. Consider a four-dimensional inflationary
model with $k{=}0$ and consider also the causal domain relevant for an
experiment beginning at $t{=}-\infty$ and ending at some arbitrary time $t{=}t_o$.
For a sufficiently small $t_o$, it is not hard to show that the number
of accessible states is given by $3\pi/G\Lambda_{initial}$. Naively, one
expects that this number of states will grow to $3\pi/G\Lambda_{final}$
as $t_o\rightarrow\infty$. However, this behavior is not universal. It is
not hard to construct examples\footnote{A particularly nice example to
work with analytically is a model
constructed as in section \ref{exactf} with a pre-potential
$W(\phi)=-4\beta_1 + 4\beta_2\cos^2(\alpha \phi)$ with $\beta_1>\beta_2$.}
where in fact the initial cosmological constant still fixes the
number of accessible states for arbitrarily large $t_o$. This behavior
arises because the apparent horizon is spacelike in these geometries. Hence in
such models, the number of degrees of freedom required to describe physical
processes throughout a given time slice grows with time while the number of
states that are accessible to experimental probing by a given physicist
remains fixed.

This discussion reminds us of the sharp contrast in
the `degrees of freedom' in the dS/CFT duality \cite{AS1} and in the
$\Lambda$-N correspondence \cite{Banks}. In the $\Lambda$-N framework,
the physics of asymptotically de Sitter universes is to be described by
a finite dimensional space of states. This dimension is precisely determined
as the number of states accessible to probing by a single observer. (The
latter is motivated in part by the conjecture of black hole complementary
\cite{STU,LS,ST}.) In contrast, in the dS/CFT context, one would
expect that a conformal field theory with a finite central charge should
have an infinite dimensional Hilbert space\footnote{Though this
infinity might perhaps be removed if one imposes, 
as described in \cite{witten}, that the conformal generators
vanish on physical states.} , and these states are all
involved in describing physical phenomena across the entire time slices.
 Further as shown above, the
central charge as a measure of number of degrees of freedom on a given
time slice need not be correlated with the number of states experimentally
accessible to observers on that slice.

A similar discrepancy between `accessible' and `available' states already
appeared in the discussion of big wide universes in section \ref{bwentropy}.
A related conceptual issue is the tension, alluded to above, between the
unitary time evolution of the bulk dS theory and the variation in
available number of degrees of freedom as manifest in the c-function.
To be concrete, consider for example the case of a dS flow foliated by
spherical spatial slices. Barring unforeseen quantum gravity
effects, one expects that the bulk operators associated with each
sphere are related to the bulk operators on any other sphere by
some quantum version of the equations of motion. In some sense
then, any information extracted from one slice should also be
available on any of the other slices and hence there is no apparent
variation in the `number of degrees of freedom.'
On the other hand, one wishes to interpret the flow between slices as
a renormalization group flow in the dual theory where `degrees of
freedom are integrated out (or in),' as indicated by the variation
of the c-function. The synthesis of these apparently
orthogonal points of view may be tied to some course-graining
scheme. However, it is logically possible that the bulk unitarity
implies that the renormalization group paradigm which was
so successfully developed for the AdS/CFT is not in fact appropriate
in a dS/CFT setting.  An alternative paradigm, or perhaps a parallel
feature, for the dS/CFT is the mapping between CFT's associated with
$I^-$ and $I^+$ in the $k{=}+1$ flows \cite{AS1,witten}. A suitable
generalization seems to describe an interesting `duality' between
different CFT's, involving a non-local mapping of operators \cite{tale2}.


\acknowledgments

The authors would like to thank Stephon Alexander, Vijay Balasubramanian,
Raphael Bousso, Marty Einhorn, David Garfinkle, Michael Green, Finn
Larsen, Robb Mann, Eric Poisson, Simon Ross, Andy Strominger and Paul Townsend
for interesting conversations. FL and RCM  were supported in part by NSERC of
Canada and Fonds FCAR du Qu\'ebec. DM was supported in part by NSF
grant PHY00-98747 and funds from Syracuse University. FL and DM
would like to express their thanks to the Perimeter Institute for
its warm hospitality during certain stages of this work. FL
would also like to thank the University of Waterloo's Department
of Physics for their hospitality during this work.
Finally we would all like to thank the Centro de Estudios Cient\'\i ficos
in Valdivia, Chile for their hospitality during the latter stages
of this work.

\appendix

\section{The many faces of de Sitter}\label{folb}

Note that we may re-cast the foliations of dS$_{n+1}$ presented in
eqs.~(\ref{metric2b}--\ref{radd}) as follows:
\be
\labell{metric2}
ds^2=-{d\rr^2\over \left({\rr^2\over\lc^2}-k\right)}
+{\rr^2\over \lc^2} d\Sigma^2_{k,n}\ .
\ee
This coordinate choice provides an interesting basis for comparison
with the following representations of dS space.

Consider the following three metrics for dS$_{n+1}$,
\be
\labell{metric1}
ds^2=-{d\rr^2\over \left(
{\rr^2\over \lc^2}-k\right)}+\left( {\rr^2\over \lc^2}-k\right)d\tau^2
+{\rr^2\over \lc^2} d\Sigma^2_{k,n-1}\ ,
\ee
where the $(n{-}1)$--dimensional metric $d\Sigma^2_{k,n-1}$ is defined
in precisely the same way as above in eq.~\reef{little}. This metric
is only really new for $k=\pm1$, since for $k{=}0$ it simply reproduces
the $k{=}0$ metric in eq.~\reef{metric2}. For $k=+1$ and
$\rho<\lc$, these are standard static coordinates on dS space where
$\rr=0$ corresponds to the position of the static observer's worldline
while $\rr=\lc$ is a cosmological horizon. In eq.~\reef{metric1},
our notation is adapted the the cosmological region (\ie $\rho>\lc$) where
this `radial' coordinate plays the role of the cosmological time, which
parametrizes the renormalization group flow. Note that for $k=\pm1$,
the scaling of the boundary metric is not homogeneous along the $\rr$-flow.

One other metric on dS space which we will consider is
\be
\labell{metric3}
ds^2=-{d\rr^2\over{\rr^2\over\lc^2}-k}
+\left({\rr^2\over\lc^2}-k\right)d\widehat\Sigma_{-k,\hat m}^2 +{\rr^2\over\lc^2}
d\widetilde\Sigma^2_{k,\tilde m}\ ,
\ee
where again the metrics $d\widehat\Sigma_{-k,\hat m}^2$ and
$d\widetilde\Sigma^2_{k,\tilde m}$ are defined in eq.~\reef{little},
and $\hat m+\tilde m=n$.
For $k{=}0$ we once again reproduce the $k{=}0$ metric in
eq.~\reef{metric2}.  For $k{=}{\pm}1$, we assume that both $\hat
m,\tilde m\ge2$.  For $k{=}{+}1$, the boundary geometry is $H^{\hat
m}{\times}S^{\tilde m}$, while for $k{=}{-}1$, we simply interchange the
hyperbolic space and the sphere. However, in the latter case, the
coordinate transformation ${\tilde \rr}^2{=}\rr^2{-}\lc^2$ puts the metric
back in the $k{=}{+}1$ form with $\hat m{\leftrightarrow}\tilde m$.
With $k{=}{+}1$, $\rr=\lc$ again corresponds to a horizon (\ie a coordinate
singularity). Finally we note that as in the previous example for $k=\pm1$,
the scaling of the boundary metric is not homogeneous as the metric evolves
in from $\rr=\infty$.

Thus with the metrics in eqs.~\reef{metric2},~\reef{metric1} and
\reef{metric3}, we have displayed dS$_{n+1}$ with a
wide variety of boundary geometries:
\be
\labell{display2}
\IR^n,\ \IR\times S^{n-1},\ \IR\times H^{n-1},\
S^{n},\ H^n,\ S^m\times H^{n-m}\ .
\ee
A dS/CFT correspondence would imply then an equivalence between
quantum gravity in dS space and a CFT on any of the above backgrounds \reef{display2}.
Of course, as discussed in section \ref{fol}, these geometries are all
related to (a portion of) $S^n$ by a singular conformal transformation.

All of the above dS metrics are maximally symmetric, {\it i.e.,} they satisfy
\be
\labell{maxi2}
R_{ijkl}={1\over\lc^2}(g_{ik}\,g_{jl}-g_{il}\,g_{jk})\ ,
\ee
which ensures that the geometry is conformally flat.
This condition also ensures the geometries are all locally dS.
One could generate additional solutions by making additional
discrete identifications of points on the spatial slices, however,
this procedure would tend to introduce null orbifold singularities
on the cosmological horizons \cite{BBM}.

However, eq.~\reef{maxi2} is an extremely restrictive condition.
If one is simply interested in solving Einstein's equations with
a negative cosmological constant
\be
R_{ij}={n\over\lc^2}g_{ij}\ ,
\ee
then the above metrics remain solutions when the spatial geometries
are replaced by arbitrary Einstein spaces. In all of the metrics
(\ref{metric2}--\ref{metric3}), one may replace any of
the $S^p$ factors (with $p{>}1$) with any space satisfying $\widetilde
R_{ab}{=}(p-1)/\lc^2\,\tilde g_{ab}$. Similarly the $H^p$ factors can be
replaced by any space satisfying $\widehat R_{ab}{=}{-}(p-1)/\lc^2\,\hat
g_{ab}$, and the $\IR^p$ factors can be replaced by any Ricci flat
solution, \ie $R_{ab}{=}0$. For example then, $S^p$ can be replaced by
a product of spheres $S^{p_1}{\times}\cdots{\times}S^{p_q}$ where
$\sum_{i=1}^qp_i{=}p$ with $p_i{>}2$ and the radii of the individual
spheres is scaled so $r_i^2{=}(p-1)/(p_i-1)\,\lc^2$. These generalized
solutions will no longer be conformally flat or locally
dS. Furthermore generically a true curvature singularity is
introduced at the minimum radius, {\it e.g.}, $R_{ijkl}R^{ijkl}$ grows
without bound as $\rr$ approaches $\lc$ or 0.

\FIGURE{\epsfig{file=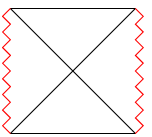, width=5cm}\caption{The
conformal diagram for negative $\mu$.}\label{negmu}}

Another simple extension of the solutions given in eq.~\reef{metric1}
comes from introducing a `charged black hole' into dS space.
The corresponding metric may be written as
\beq
\labell{metric5}
ds^{2} = - \frac{1}{H(\rho)} d\rho^{2} + H(\rho) d\tau^{2} +
\frac{\rho^{2}}{\lc^{2}}d\Sigma^{2}_{k,n-1}\ ,
\eeq
where
\beq
H(\rho) = \frac{\rho^{2}}{\lc^{2}}-k + \frac{\mu}{\rho^{n-2}} +
\frac{Q^{2}}{\rho^{2n-4}},
\eeq
and our notation is adapted to the asymptotic cosmological
regions where the `radial' variable $\rho$ appears timelike.
Of course, the full solution now contains a Maxwell field
and the electrostatic potential will have the form:
$\phi(\rho)\propto Q/\rho^{n-2}$. For $k=+1$, this reproduces
the standard Reissner-Nordstrom-dS solution.

Of course if $Q=0$, eq.~\reef{metric5} provides a vacuum solution
and for $k=+1$, this reproduces the standard Schwarzschild-dS solution.
For these vacuum solutions with $\mu$ positive, $k=+1$ is the only case
for which there {\it may be} cosmological (and black hole) event horizons.
For $k=0,-1$, the solutions have a spacelike or cosmological singularity
at $\rr=0$. If $\mu$ is negative, all of the solutions have a cosmological
horizon which separates the asymptotically dS regions from timelike singularities
at $\rr=0$, as illustrated in figure \ref{negmu}. The connection of
these solutions to the dS/CFT duality were
discussed in, \eg ref.~\cite{robb,test}. Similar comments
apply about the causal structure for the full solution with $Q\not=0$.

Interpreted as a renormalization group flow, this family of solutions \reef{metric5}
is interesting as the scaling of the $\tau$ direction is different from that for the
remaining boundary directions. However, for the vacuum solutions with $Q=0$, the
corresponding flows are trivial in that the c-function \reef{cc} is simply a fixed
constant since $\Lambda_{eff}=\Lambda$. On the other hand, with the background
Maxwell field (\ie $Q\not=0$), we have a time-dependent $\Lambda_{eff}$.
The c-function \reef{cc} for the flow under consideration has the form
\beq
\labell{eexamp}
c\sim \Lambda_{eff}^{-(n-1)/2}=\left|G^{\rho}{}_{\rho}\right|^{-(n-1)/2}\ .
\eeq
{}From eq.~\reef{metric5}, we find
\beq
\labell{endensf}
\Lambda_{eff} = -G^{\rho}{}_{\rho} =
\frac{n(n-1)}{2\lc^2} - \frac{(n-1)(n-2)Q^{2}}{2 \rho^{2(n-1)}}\ .
\eeq
Eq.~(\ref{endensf}) is a monotonically increasing function from $\rho\rightarrow +\infty$. It
is finite both on $I^{-}$ and at the cosmological horizon
$\rho=\rho_{+}$, which is the largest root for $H(\rho_+)=0$.
We find that for all possible values of $n$, $Q$ and $k$, the c-function \reef{cc}
is monotonically decreasing from $I^{-}$ to the cosmological horizon at $\rho=\rho_{+}$.
Hence this provides a nontrivial example of our c-theorem of section \ref{ct}.

\section{Numerical bouncing universe} \label{bouncers}

\FIGURE{\epsfig{file=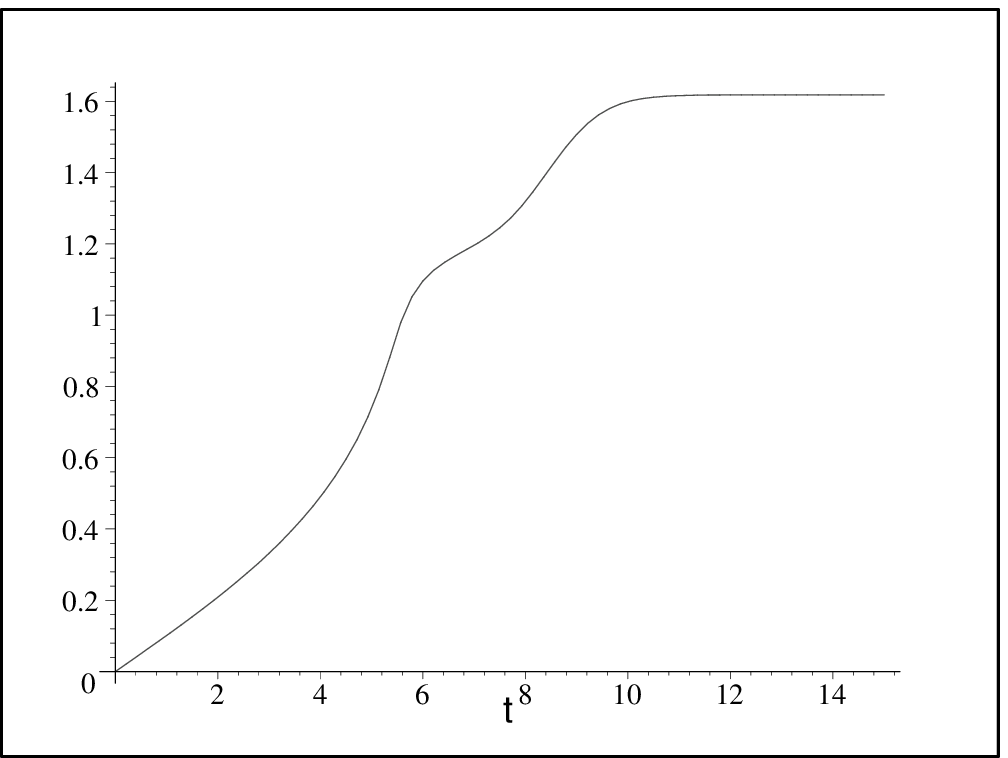, width=9cm}
\caption{\small The scalar field $\phi(t)$ increases steadily until it reaches
its maximum value at the wall of the potential where it becomes frozen.}
\label{scalar2}}
\FIGURE{\epsfig{file=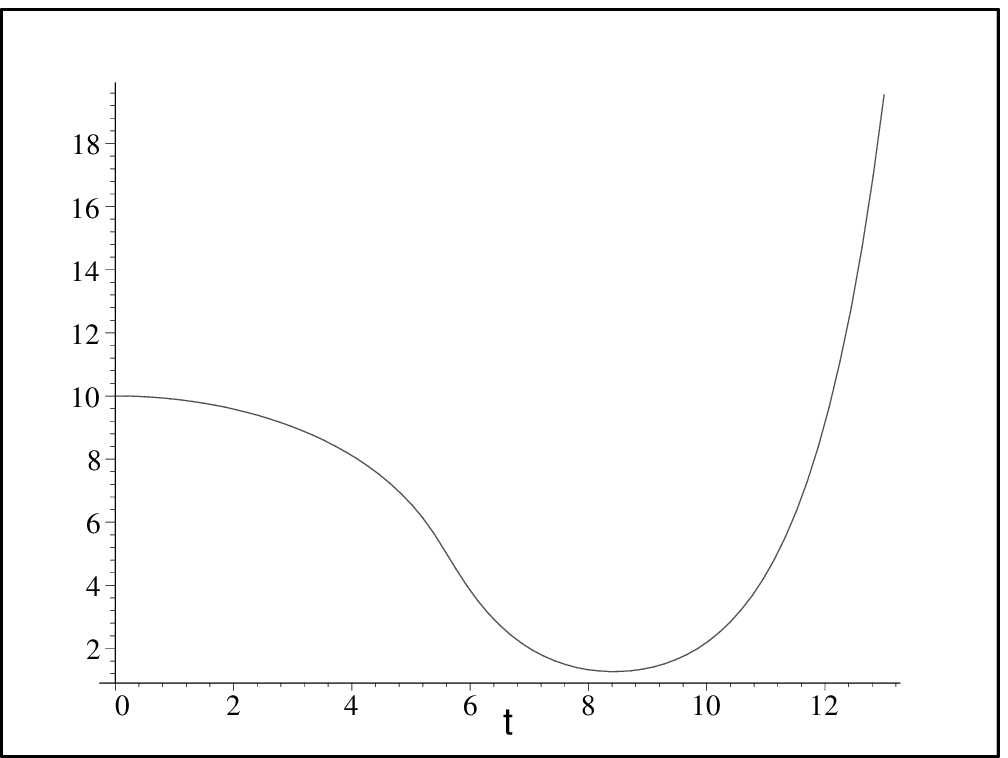, width=9cm}
\caption{\small The behavior of the scale factor is such that the universe contracts
from $t=0$ up to when the scalar field hits the wall of the potential. Then the
universe re-enters an expansion phase. The equations of motion are
symmetric when $t\rightarrow -t$ so the model of universe presented is in fact
asymptotically dS on $I^{\pm}$.}
\label{scale2}}
In section \ref{construct}, we constructed an asymptotically dS solution with
a very large matter-dominated region in the middle. The potential used there was rather
unrealistic (\ie discontinuous). To assure the skeptical reader that our results
are not an artifact of this construction, we present an analogous numerical solution here
using an everywhere smooth potential. While smooth, however, the potential
has edges which can be made arbitrarily sharp,
\begin{equation}
\labell{pot2}
V(\phi) = \frac{1}{2 L^2} \frac{\tanh w(\phi-\phi_{0})-\tanh w(\phi+\phi_{0})
+2\tanh w\phi_{0}}{\tanh w\phi_{0}}.
\end{equation}
The parameter $w$ characterizes the steepness of the potential steps and $\pm\phi_{0}$ are
approximately the values of the scalar field where the jumps occur. This potential is chosen
to be symmetric in $\phi\rightarrow-\phi$, so that we may consider a time symmetric solution.
Further the potential is constructed such that $L$ corresponds to the cosmological scale
in the asymptotic regions. That is, $\Lambda_{final}=3/L^2$ as we are working in four dimensions
($n=3$) in the following.

\FIGURE{\epsfig{file=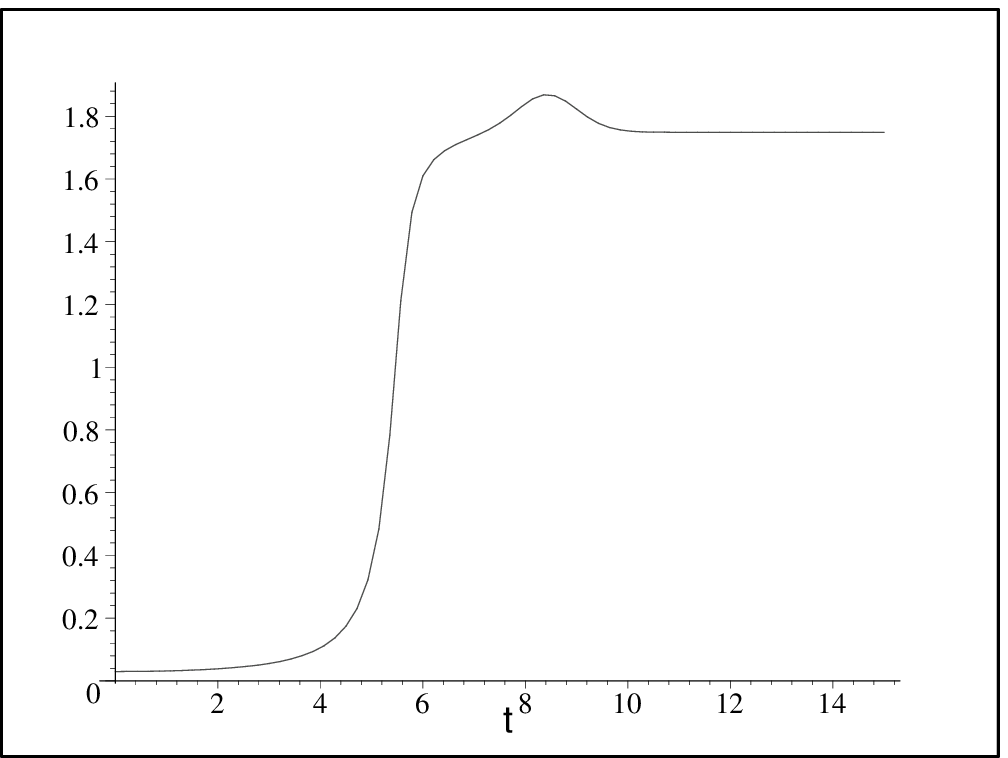, width=9cm}
\caption{\small The effective cosmological for the numerical bouncing universe is
such that the spacetime is asymptotically de Sitter in the future.} \label{cosmo2}}
Solving the second order Friedmann equation \reef{e2} numerically using Maple
yields the scalar field and a scale factor respectively shown on figures
\ref{scalar2} and \ref{scale2} given we use the initial conditions
\begin{equation}
\ a (0)= 1,\ \ \phi(0) = 0,\ \ \dot{a}(0)=0,\ \  {\rm and}\ \ \dot{\phi}(0) = 10.
\end{equation}
The parameters in eq.~\reef{pot2} are chosen such that $w=10$, $\phi_0=1$
and
\beq
\label{height1}
L = \alpha \frac{a(0)^{3}\dot{\phi}(0)}{\cosh^{\frac{3}{2}}(2\phi(0))}.
\eeq
The final parameter $\alpha$
was tuned to study various different classes of solutions. For $\alpha=1$,
the value of $L$ given by eq.~(\ref{height1}) corresponds to the maximum potential
step which the scalar could climb with the given initial conditions, according
the construction of section \ref{step}.
The figures
illustrate various aspects of the evolution for $\alpha=.9558$.
The equations of motion being left unchanged when $t\rightarrow -t$, the
flow from $t=-\infty$ to $t=0$ can be deduced simply by using the reflection
of figures \ref{scalar2} and \ref{scale2} across the $t=0$ axis. Figure
\ref{cosmo2} shows the evolution of the effective cosmological constant
$\Lambda_{eff}$. Note that asymptotically $\Lambda_{eff}$ is a constant
(as is the scalar) indicating that the evolution reaches an asymptotically
dS phase. Hence the full solution would begin in a dS phase, then enter
a matter (scalar) dominated bounce and finally return to a dS phase, just
as described for the `very tall' universe models in section \ref{construct}.

{}From examining figures \ref{scale2} and
\ref{cosmo2}, we see that $\Lambda_{eff}$ increases monotonically during
the contracting phase of the evolution and decreases in the expanding
phase, in accord with the c-theorem of section \ref{ct}.
It is interesting to note that for $\alpha > .9559$ (holding all other
parameters fixed) the scalar
continues to roll after climbing up the wall and the
resulting kinetic energy forces the scale factor to contract into
a big crunch.  On the other hand, for $\alpha < .9557$ (again holding
all other parameters fixed) the field
fails to completely climb the wall and $\phi$ returns to zero, again
resulting in a big crunch.  Only in the (approximate) range $.9557<
\alpha < .9559$ do the solutions reach an asymptotically de Sitter
regime.

\FIGURE{\epsfig{file=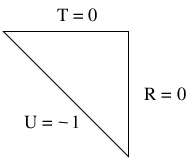,width=4cm}\caption{Conventions 
for the $R,T,U$ coordinates.}
\label{RTU}}

\section{Conformal diagram of $k=0$ de Sitter flows}
\label{flatdiagram}

In this appendix we finish the derivation of figure \ref{genflat} (b), the
conformal diagram for asymptotically de Sitter spaces with flat
spatial slices. As described in section \ref{flat}, the diagram
will consist of two regions, each of which is conformal to one
quadrant of Minkowski space and each of which has an interior that
is foliated by flat spatial hypersurfaces.  One of
these regions may be represented by a quadrant of a diamond as
shown in figure \ref{flat} (b), but this fixes part
of the conformal freedom so that the representation of the second region
is more constrained.  To find its shape it is useful to choose
explicit coordinates.
Let us begin by introducing the null coordinates $u = \tau -r$ and
$v = \tau + r$ and the corresponding $U = \tanh(u)$, $V =
\tanh(v)$. We shall also introduce $T,R$ by $U = T - R$ and $V = T
+ R$ and take the convention that our conformal diagrams are drawn
in such a way that $T,R$ appear as Cartesian coordinates.  In
particular, we take the boundaries of the triangle in figure
\ref{RTU} to lie at $T=0$, $R=0$, and $U=-1.$

Note that we retain the conformal
freedom to replace $U$
by any smooth function of $U$ in the region $U< -1$.
We may therefore use this freedom to place $I^-$ on
some convenient line, subject only to the constraint that the
angle (in the Lorentzian geometry sense) between $I^-$ and $R=0$
is unchanged.  A convenient
choice is to use the analogue of a constant $t$ line in the region
beyond $U=-1$.  That is, we define a new coordinate $u'(U) =
\tanh(2+U)$ in this region and use a line of the form $u'+v =
T_0$.

It remains only to determine the location of the left timelike
boundary of our diagram, which will represent the center of
spherical symmetry for the flat slices beyond our Cauchy horizon.
Note that this line must intersect each surface of homogeneity
orthogonally.  Thus, the shape of this boundary will be determined
if we find the surfaces of homogeneity.

\FIGURE{\epsfig{file=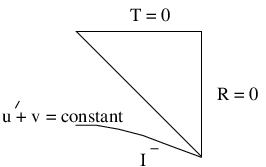,width=4cm}\caption{The 
form of the line representing $I^-$.}
\label{figlab1}}

To do so, we simply extend the coordinate $r$ to range over $[-\infty,
\infty]$ on both sides of the Cauchy horizon.
This corresponds to taking a slice all of the way across our
original higher dimensional spacetime instead of truncating the
slice at $r=0$.  For any such slice there is a translation that
reduces to $\frac{\partial}{\partial r}$ along the slice, so that
we may consider $\frac{\partial}{\partial r}$ to generate surfaces
of homogeneity in the above spacetime. The important point is
that, since $\frac{\partial}{\partial r}$ is a Killing field of
the spacetime, it must be a conformal Killing field of the
conformally re-scaled spacetime drawn in the diagram above. As a
result, it must be of the form $f(U) \frac{\partial}{\partial U} +
g(v) \frac{\partial}{\partial v}$ across the entire conformal
diagram.

Now, in the upper triangle, the expression for
$\frac{\partial}{\partial r}$ in terms of
$\frac{\partial}{\partial U}$ and  $\frac{\partial}{\partial v}$
is fixed and can be computed from the coordinate
definitions. The same function $g(v)$ must therefore give the
component of $\frac{\partial}{\partial r}$ along
$\frac{\partial}{\partial v}$ in the lower part of the diagram.
It remains only to determine the function $f(U)$ giving the
component along $\frac{\partial}{\partial U}$ for $U < -1.$  But
this is fixed by the requirement that $I^-$ be along the line $u+v
= T_0$. Since $I^-$ is a surface of homogeneity,
$\frac{\partial}{\partial r}$ must be tangent to
$\frac{\partial}{\partial u} - \frac{\partial}{\partial v}$ at
$T=-1$.  This suffices to determine $f(U)$ for $U < -1$. Without
solving these equations in detail, it is clear that the result is
simply that the left timelike boundary is a line of the form $u-v
= R_0$.

\FIGURE{\epsfig{file=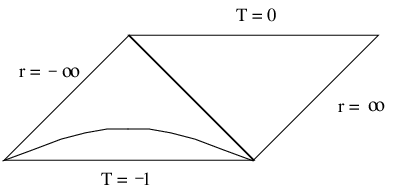,width=5cm}\caption{The
coordinate $r$ is extended to range over the real line.}
\label{figlab2}}
If one desires, one can perform a transformation (a translation in
$u$ and $v$) to the conformal frame shown below in which the
diagram has a $Z_2$ symmetry of inversion through the center.
Again, focusing arguments imply that the figure must be `taller
than it is wide,' so that the dashed congruence of light rays in
figure \ref{appflatdias} does not pass from $I^-$ to $I^+$.
\FIGURE{\epsfig{file=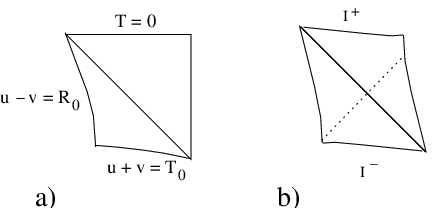, width=12cm}\caption{The diagram
for an asymptotically de Sitter spacetime with flat surfaces of homogeneity
in a) the frame described above and b) a frame where the diagram has
a reflection symmetry though the center.}\label{appflatdias}}


\end{document}